\documentclass[10pt,letterpaper]{article}
\usepackage[top=0.85in,left=2.75in,footskip=0.75in]{geometry}

\usepackage{amsmath,amssymb}
\usepackage{mathrsfs}

\usepackage{changepage}

\usepackage{textcomp,marvosym}

\usepackage{cite}

\usepackage{nameref,hyperref}

\usepackage[right]{lineno}

\usepackage[table]{xcolor}

\usepackage{array}
\usepackage{gensymb}
\usepackage{orcidlink}

\newcolumntype{+}{!{\vrule width 2pt}}

\newlength\savedwidth

\raggedright
\usepackage[aboveskip=1pt,labelfont=bf,labelsep=period,justification=raggedright,singlelinecheck=off]{caption}

\makeatletter
\renewcommand{\@biblabel}[1]{\quad#1.}
\makeatother

\usepackage{lastpage,fancyhdr,graphicx}
\usepackage{subfig,enumerate}
\usepackage{epstopdf}
\fancyheadoffset[L]{2.25in}
\fancyfootoffset[L]{2.25in}
\begin{document}

%\end{document}

\newcommand{\ub}{\mathbf{u}}
\newcommand{\Pe}{\textrm{Pe}}

\vspace*{0.2in}

\begin{flushleft}
{\Large
\textbf\newline{Natural convection in the cytoplasm: Theoretical predictions of buoyancy-driven flows inside a cell} % Please use "sentence case" for title and headings (capitalize only the first word in a title (or heading), the first word in a subtitle (or subheading), and any proper nouns).
}
\newline
\\
Nikhil Desai\orcidlink{0000-0002-0985-1384},
Weida Liao\orcidlink{0000-0002-0000-228X},
Eric Lauga\orcidlink{0000-0002-8916-2545}*,
%Name4 Surname\textsuperscript{2},
%Name5 Surname\textsuperscript{2\ddag},
%Name6 Surname\textsuperscript{2\ddag},
%Name7 Surname\textsuperscript{1,2,3*},
%with the Lorem Ipsum Consortium\textsuperscript{\textpilcrow}
\\
\bigskip
Department of Applied Mathematics and Theoretical Physics, \\ University of Cambridge, Cambridge, United Kingdom
\\
%\textbf{2} Affiliation Dept/Program/Center, Institution Name, City, State, Country
%\\
%\textbf{3} Affiliation Dept/Program/Center, Institution Name, City, State, Country
%\\
\bigskip

% Insert additional author notes using the symbols described below. Insert symbol callouts after author names as necessary.
% 
% Remove or comment out the author notes below if they aren't used.
%
% Primary Equal Contribution Note
%\Yinyang These authors contributed equally to this work.

% Additional Equal Contribution Note
% Also use this double-dagger symbol for special authorship notes, such as senior authorship.
%\ddag These authors also contributed equally to this work.

%\textcurrency Current Address: Dept/Program/Center, Institution Name, City, State, Country % change symbol to "\textcurrency a" if more than one current address note
% \textcurrency b Insert second current address 
% \textcurrency c Insert third current address


% Group/Consortium Author Note
%\textpilcrow Membership list can be found in the Acknowledgments section.

% Use the asterisk to denote corresponding authorship and provide email address in note below.
* e.lauga@damtp.cam.ac.uk

\end{flushleft}

\section*{Abstract}
The existence of temperature gradients within eukaryotic cells has been postulated as a source of natural convection in the cytoplasm, i.e.~bulk fluid motion as a result of temperature-difference-induced density gradients. Recent computations have predicted that a temperature differential of $\Delta T \approx 1$~K between the cell nucleus and the cell membrane could be strong enough to drive significant intracellular material transport. We use numerical computations and theoretical  calculations to revisit this problem in order to further understand the impact of temperature gradients on flow generation and advective transport within cells. Surprisingly, our computations yield flows that are an order of magnitude weaker than those obtained previously for the same relative size and position of the nucleus with respect to the cell membrane. To understand this discrepancy, we  develop a semi-analytical solution of the convective flow inside a model cell using a bi-spherical coordinate framework, for the case of an axisymmetric cell geometry (i.e.~when the displacement of the nucleus from the cell centre is aligned with gravity). We also calculate exact solutions for the flow when the nucleus is located concentrically inside the cell. The results from both theoretical analyses agree with our numerical results, thus providing a robust estimate of the strength of cytoplasmic natural convection and demonstrating that these are much weaker than previously predicted. Finally, we investigate the ability of the aforementioned flows to redistribute solute within a cell.  Our calculations reveal that, in all but unrealistic cases, cytoplasmic convection has a negligible contribution toward enhancing the diffusion-dominated mass transfer of cellular material.

\nolinenumbers

\section{Introduction}
The nature and behaviour of all complex lifeforms is governed by the biochemical and physical processes occurring inside their cells. Hence, cell-scale investigations of energy and mass transfer are crucial to gain a better understanding of larger scale structure and function in living organisms. Chemical reactions inside a cell are often accompanied by intra- and/or inter-cellular heat exchange~\cite{Lowell2000}. The thermal environment within and around a cell governs important tasks, such as cell-cycle regulation~\cite{Martinez1991}, cellular metabolism~\cite{Wang2020}, cell membrane function~\cite{Quinn1988}, and protein interactions~\cite{Chung2021}. This has motivated researchers to develop techniques to accurately ascertain temperatures at the scale of a single cell, leading to the recent emergence of the field of intracellular thermometry~\cite{Jaque2014, Bai2016, Uchiyama2017, Chung2021}.

 Intracellular thermometry has already contributed significantly to our understanding of the thermal landscape inside single cells. Multiple studies have  reported non-trivial temperature heterogeneities. For example, fluorescent polymeric thermometry measurements of the COS-7 cell showed that its nucleus and centrosome can be $\sim 1$~K warmer than the surrounding cytoplasm \cite{Okabe2012}. Similarly, the nucleus of living HeLa cells was also found to be $\sim 1$~K warmer than the cytoplasm, on average \cite{Hayashi2015}. Furthermore, the mitochondria of mammalian cells have been proposed as salient `hot-spots' that display elevated temperatures as compared to the rest of the cell. The higher mitochondrial temperatures result from {thermogenesis}, i.e.~the release of heat accompanying ATP synthesis. Various studies have reported mitochondrial temperatures 1--6~K higher than the cytoplasm, due to artificially induced thermogenesis, in both COS-7 and HeLa cells \cite{Gota2009, Okabe2012, Hayashi2015, Nakano2017} (see also Table 1 in Ref.~\cite{Macherel2021}).
 
While these temperature contrasts have been measured independently using a diverse array of methods~\cite{Bai2016}, there is still major controversy around the validity of these measurements~\cite{Baffou2014, Kiyonaka2015, Suzuki2015, Baffou2015, Uchiyama2017}. Mathematical models relying on macroscopic energy balance arguments have argued that the average heat generation inside a cell is so minuscule that it cannot  possibly sustain the large temperature differentials reported in intracellular thermometric experiments \cite{Baffou2014}.  Indeed, under steady operation, the temperature distribution inside the cell, $T( \mathbf{x} )$, is related to the power produced per unit volume of the cell, $\mathcal{P}$, by the heat-diffusion equation, $k \nabla^2 T = \mathcal{P}$, where $k$ is the thermal conductivity of the medium. A scaling analysis then suggests that the total power generated by the cell would cause a temperature increase on the order of $\Delta T \sim \mathcal{P} \ell_\textrm{c}^2/k$, with $\ell_\textrm{c}$ being the characteristic length scale of the cell. The typical power delivered by a cell, $\sim \mathcal{P} \ell_c^3$, is known to be on the order of $\sim 100$ pW~\cite{Loesberg1982}. Assuming the cell environment is predominantly aqueous (hence, $k \sim 1$~W~m$^{-1}$~K$^{-1}$) and the cell-size is $\ell_c \sim 10$ $\mu$m, the temperature increase resulting from this power generation would be $\sim 10^{-5}$~K,    orders of magnitude lower than the temperature differences reported by intracellular thermometry. Resolution of this apparent paradox warrants further studies, and thus, both the theory and practice behind cellular temperature measurements are fertile and fast-growing fields of research in the biological sciences.

The aforementioned discrepancy notwithstanding, the existence of finite temperature gradients within a cell could have important biophysical implications for the bulk flow of the aqueous cytoplasm. So-called ``cytoplasmic flows''  redistribute  nutrients within a cell, which in turn affects cellular functions, such as metabolism and cell division~\cite{Kamiya1981, VerchotLubicz2009}. This flow is usually actively caused by the entrainment of cytoplasm by vesicles that are driven through the cell by molecular motors along polymeric filaments. Another possible mechanism of fluid flow is based on temperature gradients within the cell causing a passive ``cytoplasmic convection''~\cite{Kessler1977, Todd1989, Howard2019}. The intuitive physical idea is that a steady temperature difference between a cell's nucleus and its membrane can cause minute changes in the density of the cytoplasm, with the warmer regions characterised by lower densities. In general, the cell geometry is such that the resulting density gradients cannot be balanced by a purely hydrostatic pressure distribution. This means that, due to their relative buoyancy, regions of higher density must settle, while those of lower density must rise, leading necessarily to natural convection within the cell~\cite{Incropera2011, Mack1968}. At the cell-scale, these circulatory flows could act as an intriguing passive complement to the active molecular-motor-driven mass transfer inside cells. Thus, to identify the most relevant mechanisms affecting intracellular material transport, it is essential to quantify the flows driven by cell-scale temperature inhomogeneities. It is important to note here that the flow just described is unavoidable for any finite temperature gradient~\cite{Mack1968}, and thus distinguishes itself from the classical Rayleigh--B\'enard convection, which occurs only in systems where temperature gradients exceed a critical threshold \cite{Chandrasekhar1961}.

While scaling arguments suggest that temperature-gradient-induced flows could be dominant in certain plant cells~\cite{Kessler1977}, further analysis is needed to better characterise the nature of these flows. Recent computational work considered the case of a model   cell with its nuclear surface warmer than the cell membrane, thus generating a temperature gradient that causes fluid flow~\cite{Howard2019}. The primary motive of that work was to investigate: (i) the influence of the temperature difference between the nucleus and the cell membrane on the flow strength inside the cell, and (ii) the ability of these flows to transport material from the nucleus to the cell membrane. Assuming both the cell and its nucleus to be spherical, it was shown, perhaps  surprisingly, that temperature-gradient-driven convection can strongly influence transport of cellular materials characterised by low diffusivities, if the nucleus is warm enough as compared to the cell membrane. However, important questions remain unexplored. For example, how would the size and position of the nucleus affect the flow? This is relevant since the size and position of the nucleus often change due to environmental and/or functional reasons~\cite{Huber2007, Gundersen2013, Chu2017}. More fundamentally, a simpler physical model of temperature-gradient-induced cytoplasmic flow would enable us to predict the flow distribution inside the cell without the need for complex computations, and thus to rationalise these predicted strong flows. In this paper, we use a combination of numerical simulations and theoretical calculations to further analyse temperature-gradient-driven flows inside model cells and provide analytical expressions for these in the limit of the nucleus being concentric with the cell membrane. Our results allow us in turn to revisit the predictions in Ref.~\cite{Howard2019} and we argue that cytoplasm flows resulting from temperature gradients are actually much weaker than previously reported.
 
In Sec.~\ref{mathMod} we introduce our system with  the mathematical model, {and} the important parameters describing the cell geometry and the material properties of the cytoplasm. We also discuss and mathematically describe the underlying physics of natural convection at the cell-scale. In Sec.~\ref{numSim}, we present results from a numerical simulation showing natural convection of the cytoplasm. Surprisingly, our computations yield flows that are an order of magnitude weaker than those obtained earlier for the same cell geometry~\cite{Howard2019}. We examine this discrepancy via theoretical calculations, in Sec.~\ref{axisymm} and~\ref{concen}. In Sec.~\ref{axisymm}, we focus on an axisymmetric arrangement of the nucleus inside the model cell, and develop a semi-analytical solution for the fluid flow in such a configuration. Next, in Sec.~\ref{concen}, we consider a concentric placement of the nucleus inside the model cell, and obtain a closed-form analytical expression for the temperature-gradient-driven flow. The solutions obtained in both these limiting cases agree perfectly with our own computations, thus giving us confidence in our numerical predictions that cytoplasm flows resulting from temperature gradients are significantly  weaker than previously reported. Finally, in Sec.~\ref{solTrans}, we perform a thorough investigation of the ability of these flows to transport solute (via advection) within a cell. Here also we reveal that, unless under unrealistic assumptions, the flows are in fact not strong enough to significantly change the distribution of cellular material, the motion of which stays largely diffusive. We finally summarise our results and provide perspectives for future investigations in Sec.~\ref{conclusion}.
 
 \section{Modelling intracellular convection}\label{mathMod}

\begin{figure}[t]
\begin{center}
\includegraphics[width=7cm]{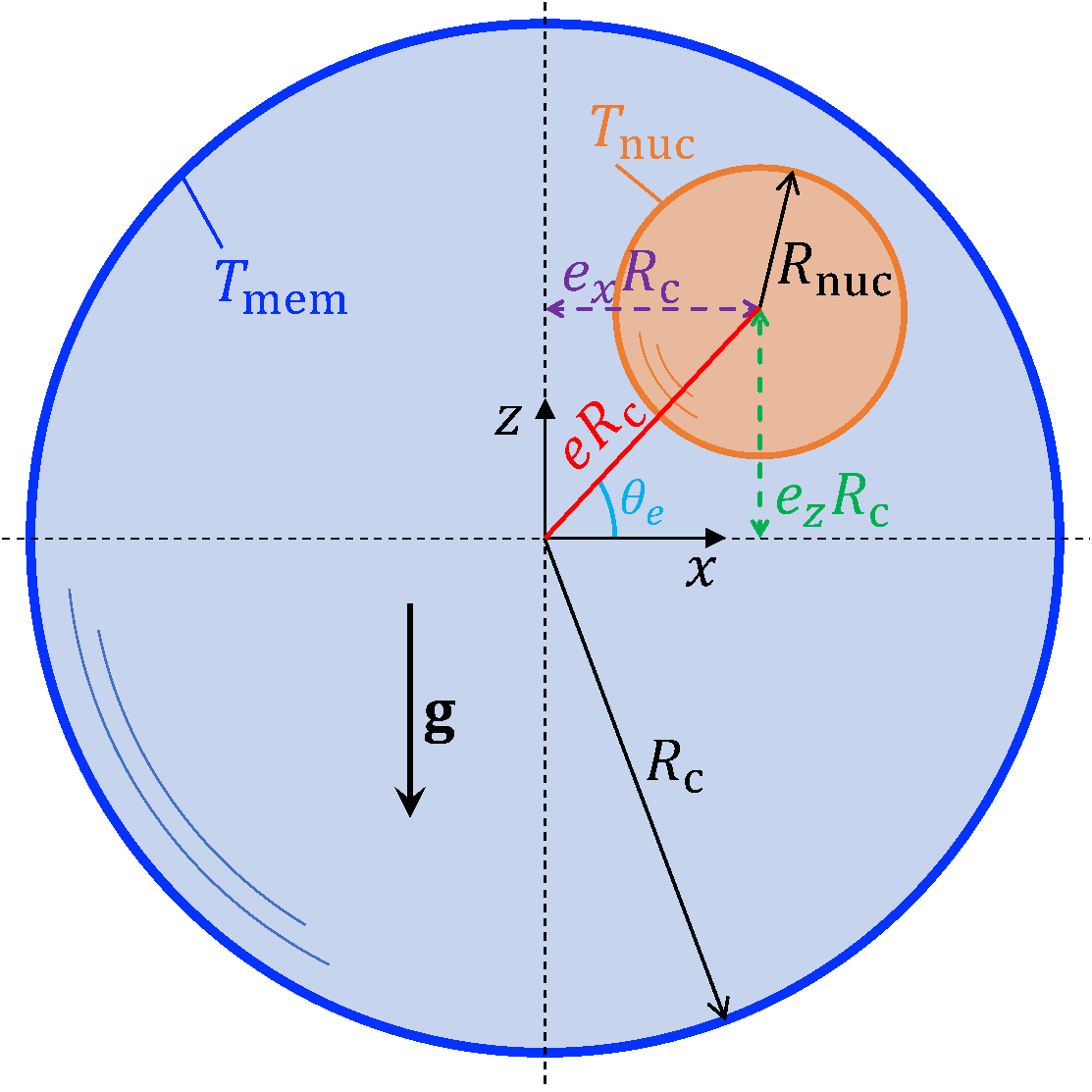}

\caption{Schematic of the nucleus (orange inner sphere) within a model  cell (light blue outer sphere). The cell membrane (thick, dark blue outline) is maintained at a constant temperature $T_\textrm{mem}$, whereas the nucleus surface is maintained at a higher temperature, $T_\textrm{nuc} > T_\textrm{mem}$. Gravity  acts in the negative $z$ direction. The Newtonian cytoplasm fills the annular region (light blue). We solve for the temperature field and fluid velocity field in the cytoplasm.}

\label{prob_schem}
\end{center}
\end{figure}

\subsection{Problem setup}\label{mathMod1}

We consider the configuration shown in Fig~\ref{prob_schem}, similar to the one addressed in Ref.~\cite{Howard2019}. Both the cell and the nucleus are modelled as stationary rigid spheres, of radius $R_\textrm{c}$ and $R_\textrm{nuc}$, respectively. A Cartesian coordinate system is affixed at the centre of the cell, with gravity acting along the negative $z$-direction. Without loss of generality, the centre of the nucleus is situated in the $y=0$ plane at a distance of $e R_\textrm{c}$ from the cell centre, at an angle $\theta_e$ from the $x$-axis. From geometry, it is clear that one must have $ e R_\textrm{c} + R_\textrm{nuc} \le R_\textrm{c}$. The quantity  {$e R_\textrm{c} \cos \theta_e = e_x R_\textrm{c}$} denotes the eccentricity, along the $x$-axis, of the nucleus centre relative to the cell centre. Similarly,  {$e R_\textrm{c} \sin \theta_e = e_z R_\textrm{c}$} denotes the nucleus eccentricity along the $z$-axis.

We assume that the cellular membrane is maintained at a constant temperature $T_\textrm{mem}$, whereas the surface of the nucleus is maintained at a constant temperature $T_\textrm{nuc}$. Following experimental evidence~\cite{Okabe2012, Chung2021}, we take the nucleus to be  warmer than the cell membrane (i.e.~$T_\textrm{nuc} > T_\textrm{mem}$). Note that while we are modelling flows due to temperature gradients between the nucleus and the cell membrane, a change in the size $\left( R_\textrm{nuc} \right)$ and position $\left( e, \theta_e \right)$ of the inner sphere can also provide an approximation of convection caused by other warm organelles inside the cell.

With cells that are typically tens of microns in size~\cite{LubyPhelps1999, Mogilner2018}, intracellular fluid flow is expected to be dominated by viscous forces and thus governed by  classical microhydrodynamics~\cite{HappelBrenner2012}. Following Ref.~\cite{Howard2019}, we model the  aqueous cytoplasm as a Newtonian viscous liquid with physical properties similar to water at a reference membrane temperature, $T_\textrm{mem}$~\cite{LubyPhelps1999, Mogilner2018}. 

We further approximate the viscosity of the cytoplasm as uniform; although viscosity is known to vary with temperature, including this effect is known to only give rise to a small correction to the flow~\cite{Leal2007}. The driving of the flow is thus solely due to the  density of the cytoplasm, which varies spatially in response to thermal gradients  and thus necessitates solving for the temperature distribution within the cell. The temperature is governed by an advection-diffusion equation, denoting a balance between heat transfer by fluid flow and by diffusion (conduction). In the next section, we formalise these statements and present the equations governing the flow and temperature fields inside the cell.

\subsection{Governing equations and boundary conditions}\label{mathMod2}

In the absence of inertia, the flow field, $\ub' \left( \mathbf{x}' \right)$ {at a position $\mathbf{x}'$ inside the cell}, is governed by the incompressibility constraint (or the continuity equation), 
\begin{equation}\label{continuity}
    \nabla' \cdot \ub' = 0,
\end{equation}
and the Stokes equations,
\begin{equation}\label{Stokes_natural_conv}
    -\nabla' p' + \eta \nabla'^2 \ub' + \rho(T') \mathbf{g} = \mathbf{0},
\end{equation}
where the primes  {signify} dimensional variables. 

In the above equation, $p'$ is the pressure in the fluid, $\eta$ is the  constant dynamic viscosity, $\mathbf{g} = -g \mathbf{i}_z$ is the acceleration due to gravity ($\mathbf{i}_z$ is the unit vector in the $z$-direction), and $T'$ is the temperature in the cell. We denote the temperature-dependent density of the fluid by $\rho(T')$. For sufficiently small temperature differences $T'-T_\textrm{mem}$, we model this temperature-dependence with a standard linear relationship,

\begin{equation}\label{rho_T}
    \rho(T') = \rho_0 \left[ 1 + \beta(T_\textrm{mem} - T') \right],
\end{equation}
where $\beta$ is the thermal expansion coefficient, with units of K$^{-1}$.  
It is important to note here that we are employing the incompressibility constraint, eqn.~\eqref{continuity}, yet allowing the density to vary with temperature. {This is a classical (and validated) method in hydrodynamics called the Boussinesq approximation~\cite{Leal2007}. For modest changes in temperature, we can neglect the temperature dependence of material properties everywhere except in the gravitational force term in eqn.~\eqref{Stokes_natural_conv},  when this buoyancy term is the driving mechanism for the flow (as in the present problem). This provides a first approximation to the fluid flow, which should be quantitatively accurate for temperature differences less than $\sim 10$~K--$20$~K~\cite{Leal2007}.} Mathematically, this may be formalised via a perturbation expansion of the full governing equations in the limit of small temperature changes~\cite{Leal2007}.

The governing equations, eqns.~\eqref{continuity} and \eqref{Stokes_natural_conv}, must be complemented by appropriate boundary conditions for the velocity. Since we are modelling the nucleus and the cell as stationary, rigid spheres, the fluid velocity must vanish at both these surfaces, 
i.e.
\begin{align}\label{u_BCs}
    \left. \ub' \right|_\textrm{nuc} &= \mathbf{0}, \\ \nonumber
    \left. \ub' \right|_\textrm{mem} &= \mathbf{0},
\end{align}
where the sub-scripts `nuc' and `mem' denote the cell nucleus and membrane, respectively. It is clear from eqn.~\eqref{Stokes_natural_conv} that solving for the fluid flow requires knowledge of the temperature. The steady-state temperature field, $T'\left( \mathbf{x}' \right)$, is governed by an advection-diffusion equation,
\begin{equation}\label{advDiff_T}
   \ub' \cdot \nabla' T' = \alpha \nabla'^2 T',
\end{equation}
subject to the boundary conditions,
\begin{align}\label{T_BCs}
    \left. T' \right|_\textrm{nuc} &= T_\textrm{nuc}, \\ \nonumber
    \left. T' \right|_\textrm{mem} &= T_\textrm{mem}.
\end{align}
In eqn.~\eqref{advDiff_T}, $\alpha$ is the constant thermal diffusivity of the cytoplasm (assumed to be water in the present study).

{It is important to note that for the geometry described in Fig~\ref{prob_schem}, eqns.~\eqref{continuity} to \eqref{T_BCs} cannot admit a quiescent solution {($\ub'=\mathbf{0}$)} as long as there exists a temperature contrast between the two surfaces. For any $T_\textrm{nuc} \ne T_\textrm{mem}$, the temperature distribution $T' \left( \mathbf{x}' \right)$ will be non-uniform and the resulting temperature gradients (and hence the density gradients) will not be aligned with gravity inside the cell. Thus, no hydrostatic pressure distribution can be found that balances the gravitational forcing everywhere, and the system cannot stay in equilibrium without fluid flow. This is fundamentally different from the classical Rayleigh--B\'enard convection, where gravity is parallel to an imposed temperature gradient and  fluid flow emerges as a result of an instability above a critical imposed temperature gradient \cite{Chandrasekhar1961}.}

We next render the equations dimensionless; using the  characteristic temperature  difference $\Delta T = T_\textrm{nuc} - T_\textrm{mem}$,  we may define the   reference scales for length, velocity and pressure, respectively, as
\begin{equation}\label{scales}
    \ell_\textrm{ref} = R_\textrm{c},\quad u_\textrm{ref} = \frac{\rho_0 \beta \Delta T g \ell_\textrm{ref}^2}{\eta}, \quad  p_\textrm{ref}=\frac{\eta u_\textrm{ref}}{\ell_\textrm{ref}}.
\end{equation}
 We also define a re-scaled temperature,
\begin{equation}\label{theta}
    \Theta = \frac{T' - T_\textrm{mem}}{T_\textrm{nuc} - T_\textrm{mem}}.
\end{equation}

\begin{table}[t]
\begin{tabular}{ p{0.2\linewidth} | p{0.41\linewidth} | p{0.17\linewidth} | p{0.175\linewidth} }
 {Parameter}    &  {Description} &  {Typical value} &  {Units}    \\
\hline
$g$ & gravitational force per unit mass & 9.8 & m s$^{-2}$ \\
$\beta$ & thermal expansion coefficient & $3.61 \times 10^{-4}$ & K$^{-1}$ \\
$\Delta T = T_\textrm{nuc} - T_\textrm{mem}$ & temperature difference between the nuclear surface and the cell membrane & 0.01--10 & K \\
$R_\textrm{c}$ & radius of the cell & 10 & $\mu$m   \\
$R_\textrm{nuc}$ & radius of the nucleus & 2--7 & $\mu$m   \\
$\rho_0$ & density of the cytoplasm at 310~K & 993.38 & kg m$^{-3}$ \\
$\eta$ & viscosity of the cytoplasm & $6.917 \times 10^{-4}$ & kg m$^{-1}$ s$^{-1}$ \\
$\alpha$ & thermal diffusivity of the cytoplasm & $1.51 \times 10^{5}$ & $\mu$m$^2$ s$^{-1}$ \\
$D$ & solute diffusivity & 0.01--100 & $\mu$m$^2$ s$^{-1}$ \\
\hline
$u_\textrm{ref} = \frac{\rho_0 \beta \Delta T g R^2_\textrm{c}}{\eta}$ & reference velocity scale & $0.5$ 
& $\mu$m s$^{-1}$ \\
$\Pe_\textrm{t}$ = $\frac{u_\textrm{ref}R_\textrm{c}}{\alpha}$ & thermal P\'eclet number & $3 \times 10^{-5}$ 
& {\it dimensionless} \\
$\Pe_\textrm{s}$ = $\frac{u_\textrm{ref}R_\textrm{c}}{D}$ & `solutal' P\'eclet number & $0.05$--$500$  
& {\it dimensionless}
\end{tabular}
\caption{Typical values of the physical parameters (above line), the reference velocity scale $u_\textrm{ref}$, and the dimensionless groups (last two rows) governing the intracellular fluid flow, along with the heat and mass transfer.}

\label{table_of_values}
\end{table}

 Using eqns.~\eqref{scales} and \eqref{theta} in the equation for the flow, eqn.~\eqref{Stokes_natural_conv}, yields  now in dimensionless form (without primes)
\begin{equation}\label{Stokes_natural_conv_ND}
    -\nabla P + \nabla^2 \ub + \Theta \mathbf{i}_z = \mathbf{0}, \qquad \nabla \cdot \ub = 0,
\end{equation}
 subject to the dimensionless boundary conditions
 \begin{align}\label{u_BCs_dl}
    \left. \ub \right|_\textrm{nuc} &= \mathbf{0}, \\ \nonumber
    \left. \ub \right|_\textrm{mem} &= \mathbf{0}.
\end{align}
The pressure $P$ in eqn.~\eqref{Stokes_natural_conv_ND} is now a modified dynamic pressure, given in dimensionless form by $P = p + \left( \beta \Delta T \right)^{-1} \mathbf{i}_z \cdot \mathbf{x}$. 
Similarly, we can re-write the thermal problem, eqn.~\eqref{advDiff_T}, in terms of the normalised temperature $\Theta \left( \mathbf{x} \right)$ as

\begin{equation}\label{advDiff_T_ND}
    \ub \cdot \nabla \Theta = \frac{1}{\Pe_\textrm{t}}\nabla^2 \Theta,
\end{equation}
with the normalised boundary conditions,
\begin{align}\label{T_BCs_ND}
    \left. \Theta \right|_\textrm{nuc} &= 1, \\ \nonumber
    \left. \Theta \right|_\textrm{mem} &= 0.
\end{align}
Importantly, in eqn.~\eqref{advDiff_T_ND}, $\Pe_\textrm{t} \equiv u_\textrm{ref}R_\textrm{c}/\alpha$ is the (dimensionless) thermal P\'eclet number, a ratio of the rate of heat transfer by fluid flow to the rate of heat transfer by diffusion.

The typical values of the parameters used for calculating the reference scales in our problem are given in Table~\ref{table_of_values}, where, just like Ref.~\cite{Howard2019}, we have assumed that the cytoplasm shares the material properties of water at temperature $T_\textrm{mem}=310$~K ($\approx 37\degree$C). This is a reasonable assumption for both the physical~\cite{LubyPhelps1999, Kalwarczyk2011} and thermal~\cite{KyooPark2013, Inomata2023} properties of the cytoplasm. We note, however, that in some instances the effective viscosity of the cytoplasm can be higher  than that of water, due to a high concentration of macromolecules~\cite{LubyPhelps1999}; we will discuss the implications of this disparity in Sec.~\ref{est_flow_BiSp} and argue that it does not alter our central result. The size (radius) of the cell in our study corresponds roughly to the typical COS-7~\cite{Okabe2012} or HeLa~\cite{Hayashi2015} cells used in intracellular thermometry studies, while we consider a wide range of sizes and positions of the nucleus, to reflect the diversity in the nucleus's placement inside biological cells, depending on the cell type and the stage of the cell cycle~\cite{Huber2007, Gundersen2013}. Here too, we note that the main conclusions of this study remain independent of the details of nuclear size and position inside the cell (see Appendix~\ref{infl_geom}).

A quantitative description of natural convection in the model cell requires us to solve eqns.~\eqref{Stokes_natural_conv_ND}--\eqref{T_BCs_ND} for $\Theta$ and $\ub$. For a prescribed thermal P\'eclet number (set by the material properties of the cytoplasm), the solution depends only on the size and position of the nucleus. These quantities are captured in three dimensionless numbers: $\kappa = R_\textrm{nuc}/R_\textrm{c}$, the ratio of the nucleus's radius to that of the cell, and $\left( e_x, e_z \right)$, the eccentricity of the nucleus centre along the $x$- and $z$-axes.

Depending on the specific values of $\left( e_x, e_z \right)$ and $\Pe_\textrm{t}$, one can solve eqns.~\eqref{Stokes_natural_conv_ND}--\eqref{T_BCs_ND} using full numerical simulations, semi-analytical methods, or even analytically to obtain an explicit expression for the temperature and flow, $\left( \Theta, \ub \right)$. In following sections, we describe solutions obtained using each of these methods to provide rigorous estimates of the flow inside the cell.

\section{Temperature and flow inside the cell}\label{numSim}

\subsection{Intracellular convective flows: comparison with previous work}

\begin{figure}[t]

\includegraphics[width=\linewidth]{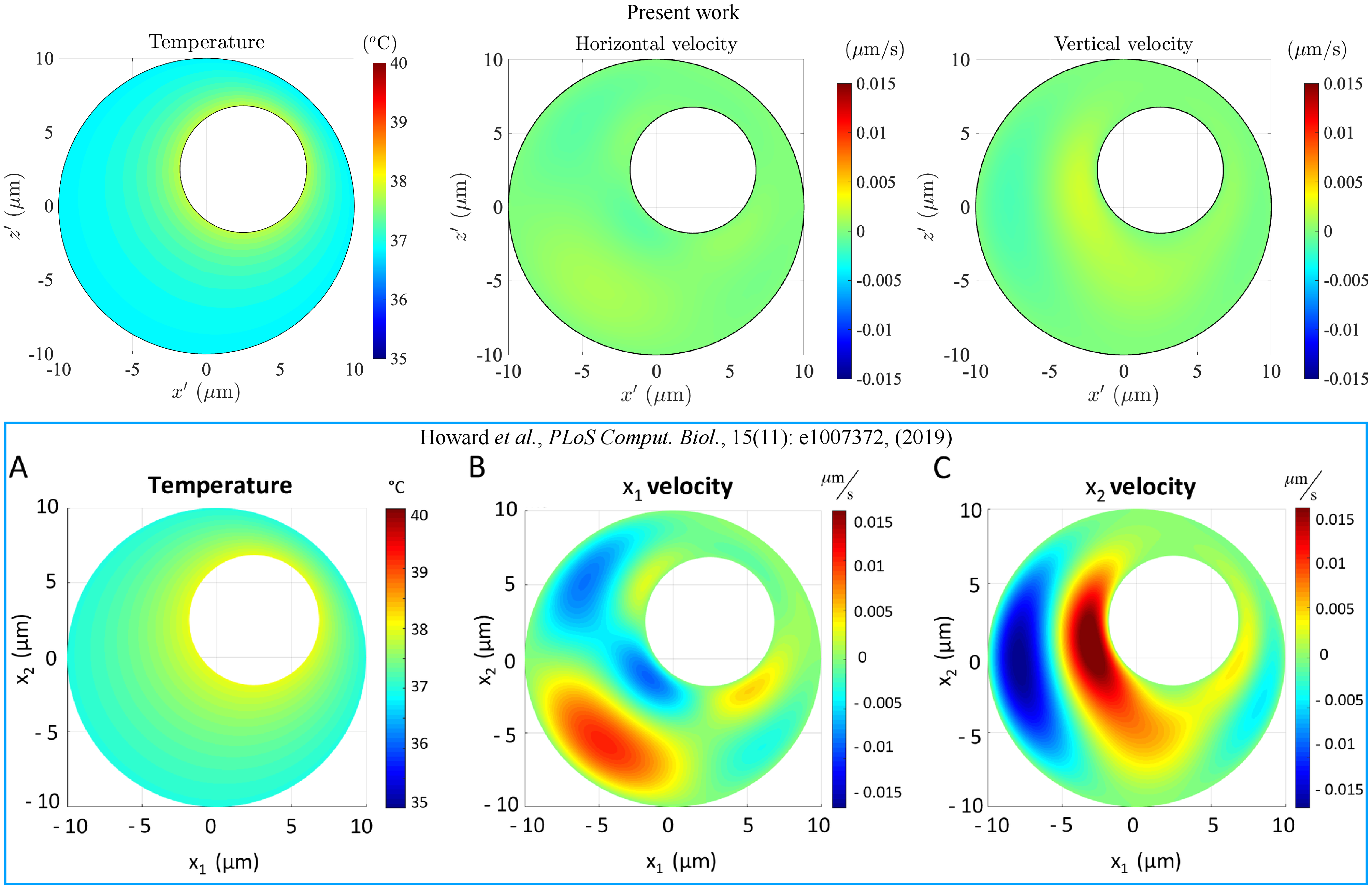}

\caption{Top row, left to right: the temperature, the horizontal velocity ($u'_x$) and the vertical velocity ($u'_z$) at the $y=0$ mid-plane of the cell, computed by simulating the advection-diffusion equation {(eqn.~\eqref{advDiff_T_ND})} for the temperature, and the Stokes equations (eqn.~\eqref{Stokes_natural_conv_ND}) for flow field. Bottom row: the corresponding fields as computed numerically in Ref.~\cite{Howard2019} for the same set of physical parameters. In our results, the spatial variation in flow velocity is barely discernible when plotted using the colour scale of Ref.~\cite{Howard2019}, as the velocity magnitudes we obtain are significantly lower.}\label{vel_contours_comp_scales}
\end{figure}

 We first present dimensional results of the temperature and flow fields inside the cell, to gain intuition about the strength of the temperature-gradient-driven flow. We show in the top row of Fig~\ref{vel_contours_comp_scales} one specific solution, for a radius ratio $\kappa = 0.43$,  eccentricity values $e_x=e_z=0.25$, and a nucleus-to-membrane temperature difference of $\Delta T = 1$~K, evaluated numerically using the finite-elements-based software COMSOL; this is the benchmark geometry used in Ref.~\cite{Howard2019}. The values of the other physical parameters are stated in Table~\ref{table_of_values} and are the same as in Ref.~\cite{Howard2019}, allowing us to compare our results directly. The results of Ref.~\cite{Howard2019} are reproduced in the bottom row of Fig 2. For the flow field, we plot the horizontal (i.e.~perpendicular to gravity) velocity $u'_x$ and the vertical (i.e.~upward) velocity $u'_z$ at the $y=0$ mid-plane of the cell. Since this is a plane of symmetry, the velocity component normal to it, i.e.~$u'_y$, is identically zero.

While the temperature distribution in Ref.~\cite{Howard2019} is only slightly different from our own simulations, the intracellular flows in that work are much stronger (typically by {around} one order of magnitude). It is clear, intuitively,  that the small difference in the temperature profiles is not sufficient to explain the ten-fold mismatch in velocity magnitudes. 

To resolve this apparent conflict and verify the accuracy of our own simulations, we will solve, in Secs.~\ref{axisymm} and  \ref{concen}, for the temperature and flow distributions in geometrically simpler domains, for which we can also pursue alternate solution methodologies. {Towards this, we first need to examine the temperature field inside the cell.}  {We will show that fluid flow is ineffective in facilitating temperature redistribution, which will allow us to simplify the governing equations further. This enables the use of semi-analytical and analytical techniques to solve the problem, validate our numerical simulations, and confirm the discrepancy between our results and those of Ref.~\cite{Howard2019}.}

\subsection{Relative influence of diffusion and advection in intracellular heat transfer}

\begin{figure}[t]

\includegraphics[width=\linewidth]{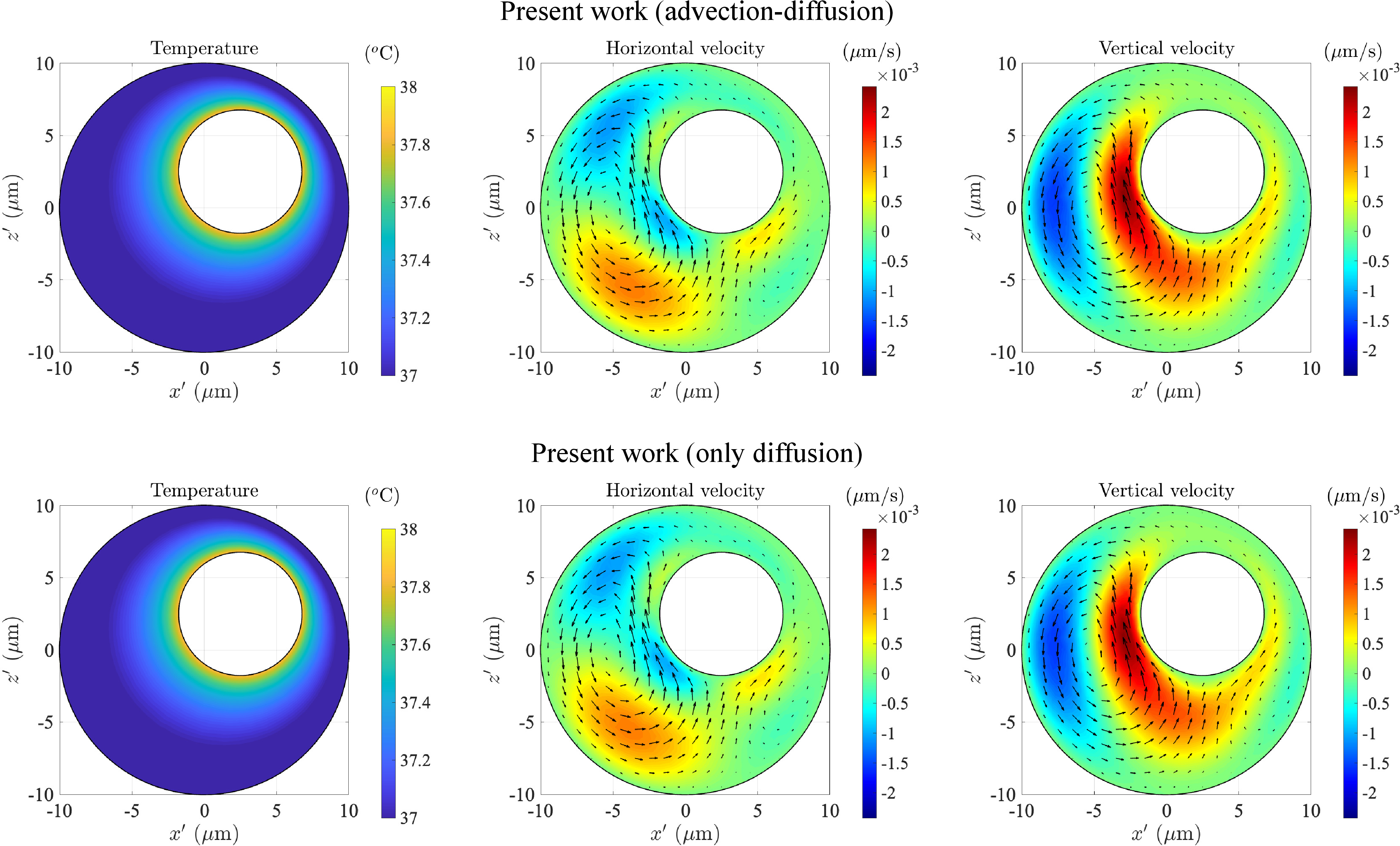}

\caption{Top row, left to right: the temperature, the horizontal velocity ($u'_x$) and the vertical velocity ($u'_z$) at the $y=0$ mid-plane of the cell, computed by simulating the advection-diffusion equation (eqn.~\eqref{advDiff_T_ND}) for the temperature, and the Stokes equations (eqn.~\eqref{Stokes_natural_conv_ND}) for flow field. Bottom row: temperature and flow fields, obtained by simulating the diffusion equation (eqn.~\eqref{lap_T_ND}) for the temperature, and the Stokes equations (eqn.~\eqref{Stokes_natural_conv_ND}) for the flow field. The arrows in the velocity contour plots denote velocity vectors, with arrow lengths proportional to the magnitude of the local velocity.}\label{vel_contours_AD_vs_oD}
\end{figure}

We show two different  simulation results in Fig~\ref{vel_contours_AD_vs_oD}. The top row corresponds to simulations where we solve the full advection-diffusion equation (eqn.~\eqref{advDiff_T_ND}) in conjunction with the hydrodynamic problem;   these are the same results   as in the top row of Fig~\ref{vel_contours_comp_scales} but  plotted with a different colour scale in order to highlight the flow variations inside the cell. 
 Given the  relation between density and temperature in  eqn.~\eqref{rho_T},  clearly the  fluid near the nucleus is lighter, while the fluid near the cell membrane is heavier.  Thus, as expected from physical intuition, the warmer fluid rises under the influence of gravity, while the colder fluid settles, leading to cell-scale circulation in the bulk. Note that we focus predominantly on the flow in the vertical plane $y'=0$, which is also a plane of symmetry for the cell, since the flows are strongest and most intuitively visualised here; we comment briefly in Appendix~\ref{y_velocity} on the nature of horizontal flow along the $y'$ direction.

In the bottom row of Fig~\ref{vel_contours_AD_vs_oD}, on the other hand, we show results where advective heat transfer is neglected and the thermal problem is governed solely by Laplace's equation,
\begin{equation}\label{lap_T_ND}
    \nabla^2 \Theta = 0,
\end{equation}
along with the boundary conditions  (eqn.~\eqref{T_BCs_ND}). The latter choice of simulations is motivated by the particularly low value of the thermal P\'eclet number in the present problem, $\Pe_\textrm{t} \sim 10^{-5}$ (see Table \ref{table_of_values}),  hinting that heat transport is likely dominated by diffusion. 

Comparing the two rows of Fig~\ref{vel_contours_AD_vs_oD}, we see that the temperature and velocity fields are practically identical in the simulations with and without temperature advection by the flow. The transfer of heat  due to the convective flow is thus essentially negligible in comparison to its rapid diffusion throughout the cytoplasm. Interestingly, as a result, the temperature field is symmetric around the line joining the nucleus and the cell centres, indicative of isotropic diffusion from a warmer to a colder surface, with no breaking of symmetry due to the convective flow.

\subsection{Weak advective effects result in one-way coupling between temperature and flow fields}

 The fact that advection appears to play a negligible role in influencing the intracellular temperature distribution offers a great simplification in terms of solving the thermal and hydrodynamic problems. We do not need solve the full advection-diffusion equation, eqn.~\eqref{advDiff_T_ND}, for the dimensionless temperature $\Theta$, but can instead consider just the diffusion equation given by eqn.~\eqref{lap_T_ND}. It then follows  that eqns.~\eqref{Stokes_natural_conv_ND} and \eqref{lap_T_ND} are only one-way coupled, with the flow being affected by the temperature, but not vice-versa. One can thus first obtain $\Theta$ and then solve the Stokes flow problem driven by the non-homogeneous temperature field $\Theta \left(\mathbf{x}\right)$.

 Equipped with the above understanding, we are now in a position to explore further the order-of-magnitude disparity apparent in Fig~\ref{vel_contours_comp_scales}. In the subsequent sections, we compare our numerical simulation results with solutions obtained through other solution strategies that can be employed in slightly different geometries. We focus first on the geometric limit $e_x = 0$, i.e.~an axisymmetric placement of the nucleus within the cell. Next, we move to an even simpler configuration, wherein the nucleus is concentric with the cell, i.e.~$\left( e_x = 0, e_z = 0 \right)$. We will show that solutions in these two limits, obtained via classical techniques, match very well with our simulation results. This will thus confirm the accuracy of our numerical simulations and establish that temperature-gradient-driven intracellular flows are indeed significantly weaker than predictions made in earlier studies.
 
 \section{Temperature and flow in eccentric but axisymmetric configurations}\label{axisymm}

\begin{figure}[t]
\begin{center}

\includegraphics[width=\linewidth]{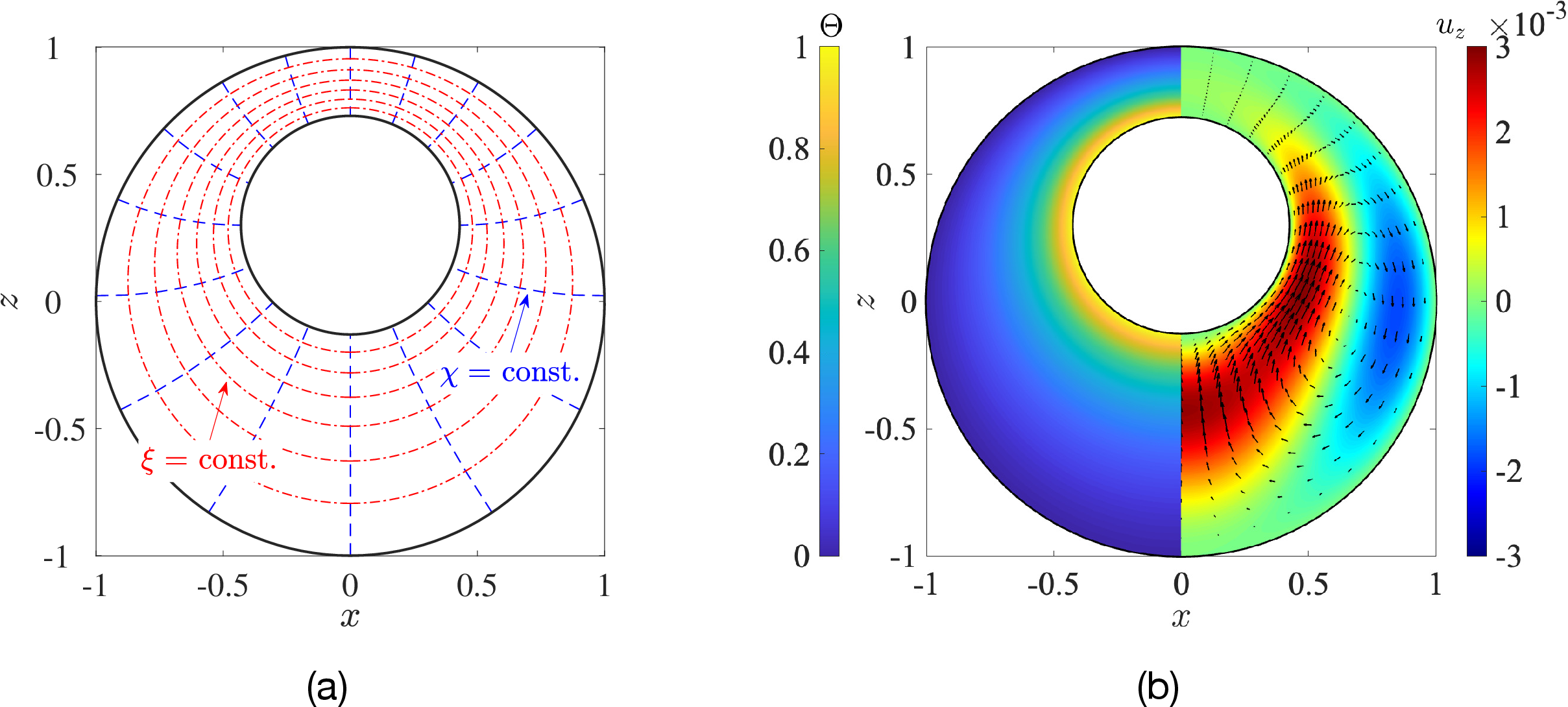}

\caption{Flow in axisymmetric cell-nucleus configuration. (a) Depiction of the bi-spherical coordinate grid. (b) An example of dimensionless results  obtained via the bi-spherical coordinate calculations. Left:  normalised temperature distribution, $\Theta$; right: vertical flow, $u_z$. The geometric parameters are: radius ratio $\kappa = 0.43$,  eccentricities $e_x=0$ and $e_z=0.3$.}
\label{BiSp_grid}
\end{center}
\end{figure}

In this section, we consider an axisymmetric arrangement of the nucleus within the cell, i.e.~when the line joining the nucleus and cell {centres}  is directed along (or opposite to) the gravity vector.

\begin{figure}[t]
\begin{center}

\includegraphics[width=\linewidth]{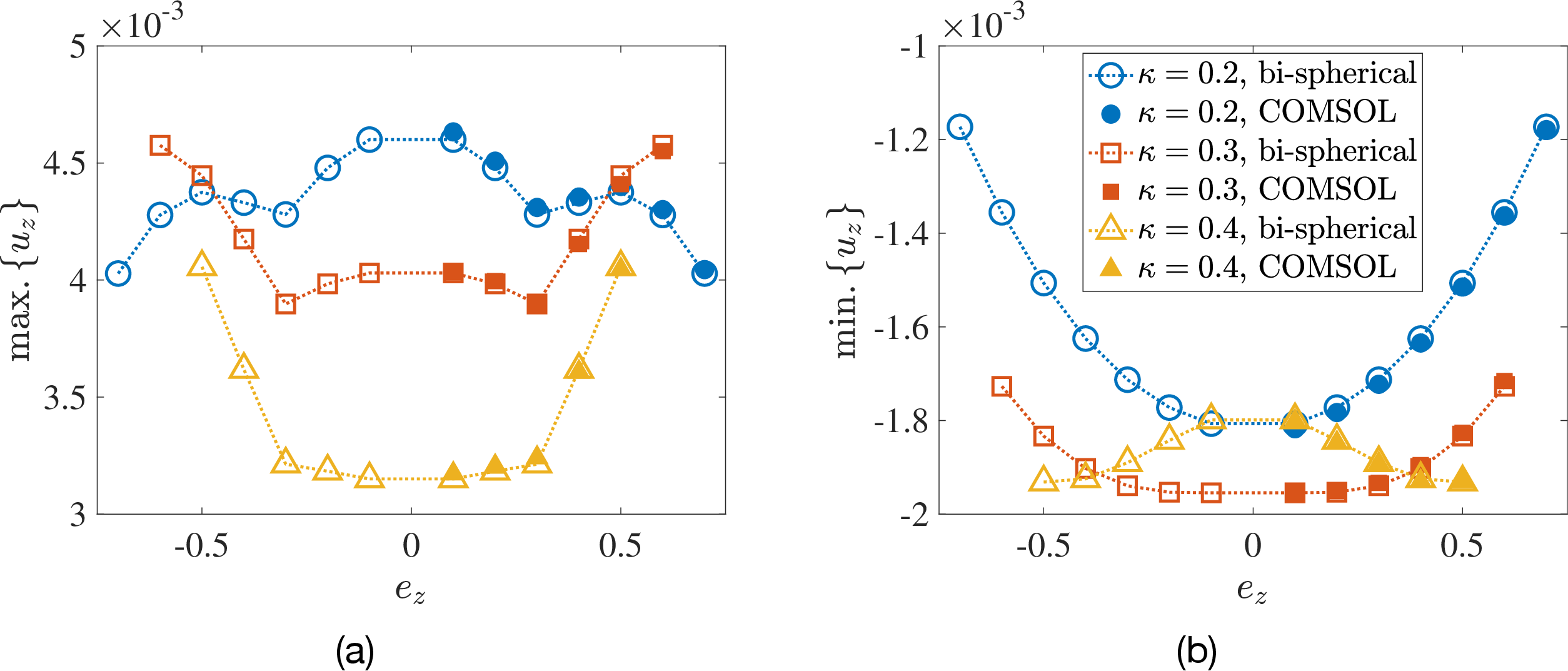}

\caption{Dimensionless values of the (a) maximum upwelling velocity, max.\{$u_z$\}, and (b) minimum downwelling velocity, min.\{$u_z$\}, as a function of the vertical eccentricity $e_z$ for $\kappa = 0.2, 0.3, 0.4$, obtained using the semi-analytical  bi-spherical coordinates calculations (empty symbols) and via COMSOL simulations (filled symbols).}
\label{ecc_val_comsol_3D}
\end{center}
\end{figure}

\subsection{Outline of solution in axisymmetric geometry}
 
An important implication of analysis in axisymmetric geometry is the symmetry of the flow field around the  axis between the nucleus and the centre of the  cell. 
We may exploit this symmetry to solve eqns.~\eqref{Stokes_natural_conv_ND} and \eqref{lap_T_ND} semi-analytically in a bi-spherical coordinate system, $\left( \xi, \chi, \phi \right)$, shown in Fig~\ref{BiSp_grid}(a), where the nuclear  {membrane} and the cell membrane can be represented by distinct coordinate surfaces~\cite{HappelBrenner2012}. Axisymmetry allows  the flow field to be  represented in terms of a single  {Stokes streamfunction}, $\Psi^\textrm{b}\left( \xi, \chi \right)$, which can be expressed as a linear superposition of harmonic functions~\cite{HappelBrenner2012}. The technical aspects of the solution are detailed in Appendix~\ref{bisp_calc}. A sample solution in the axisymmetric case is shown in Fig~\ref{BiSp_grid}(b), where the $z$-axis is the axis of symmetry and the geometric parameters are $\kappa = 0.43, e_x = 0, e_z = 0.3$ (left: temperature; right: vertical velocity). We can also solve the axisymmetric problem using finite-element COMSOL simulations. In Fig~\ref{ecc_val_comsol_3D}, we compare the results obtained from the two different methods, by plotting the dimensionless maximum $\left( \textrm{max}. \left\{ u_z \right\} \right)$ and minimum $\left( \textrm{min}. \left\{ u_z \right\} \right)$ vertical velocities inside the cell, as a function of the eccentricity $e_z$. We provide further validation of the temperature and velocity fields in Fig~\ref{BiSp_vs_COMSOL} in Appendix~\ref{bisp_calc}. The excellent matching between the two sets of results validates the non-axisymmetric simulations from Sec.~\ref{numSim}.

\begin{figure}[t]
\begin{center}

\includegraphics[width=\linewidth]{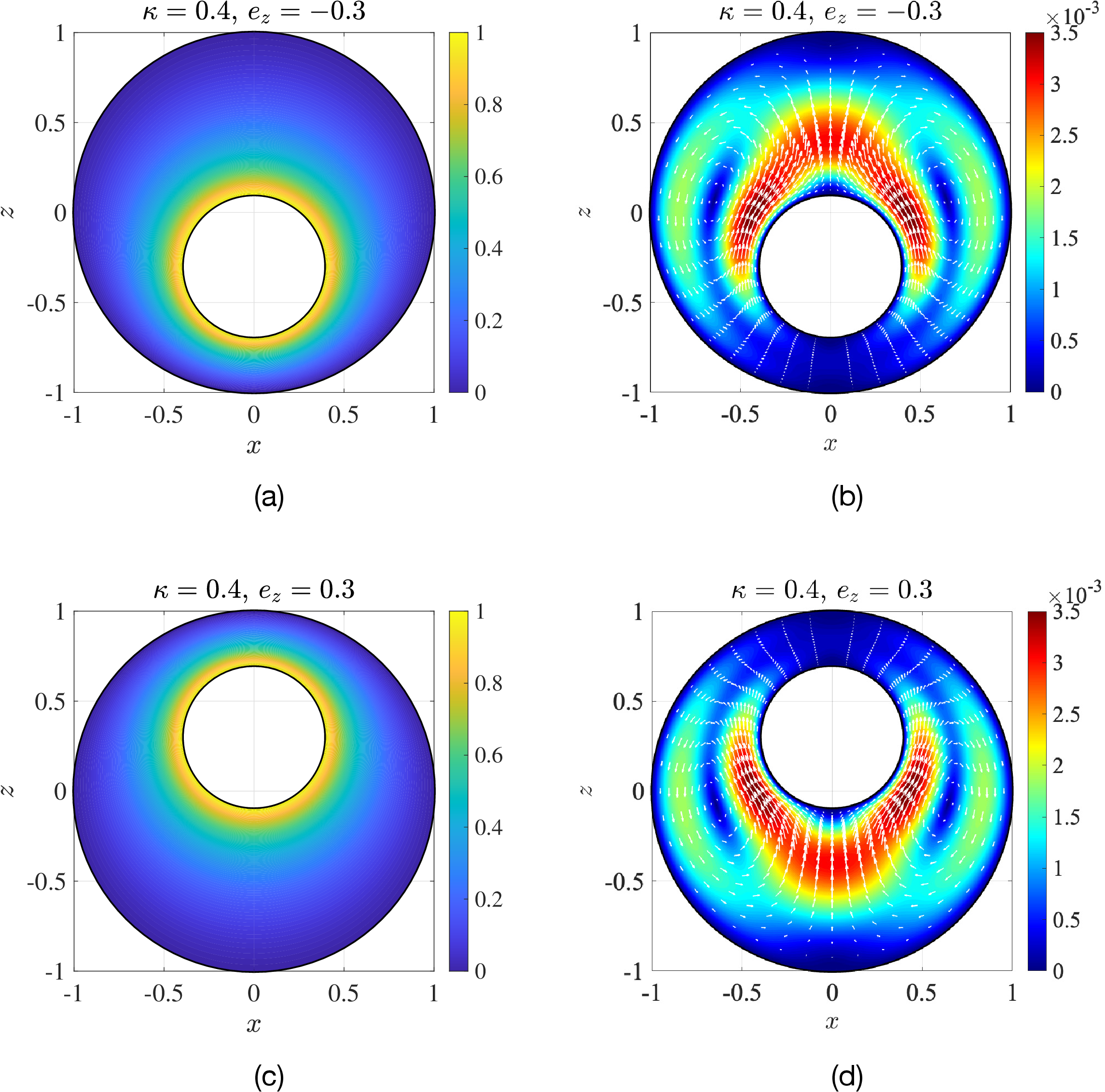}

\caption{Symmetry of the temperature $\Theta$ (panels (a) and (c)) and the flow magnitude $\left| \ub \right|$ (panels (b) and (d)) inside the cell with respect to a change in sign of the nucleus's vertical eccentricity $e_z$. The arrows are velocity vectors with lengths proportional to the velocity magnitude shown in the contour plots.}
\label{e_sign}
\end{center}
\end{figure}

\subsection{Estimating the flow strength}\label{est_flow_BiSp}

The typical magnitude of the intracellular flow velocities can be estimated by multiplying the dimensionless results in Fig~\ref{ecc_val_comsol_3D} by the reference velocity $u_\textrm{ref} \approx 0.5$~$\mu$m~s$^{-1}$ used to non-dimensionalise the equations (see Table \ref{table_of_values}); this yields a maximum velocity of $\sim 10^{-3}$~$\mu$m~s$^{-1}$, similar to what was displayed in Fig~\ref{vel_contours_AD_vs_oD}. Thus, the axisymmetric solutions provide an independent measure of the order of magnitude of cellular flows predicted by our numerical simulations in Sec.~\ref{numSim} and confirm the ten-fold discrepancy with the results of Ref.~\cite{Howard2019}.  We note here that the value of the reference velocity $u_\textrm{ref} \approx 0.5$~$\mu$m~s$^{-1}$ is based on the assumption that the cytoplasm viscosity is the same as that of water at 37$\degree$C, i.e. $\eta \approx 7 \times 10^{-4}$~kg~m$^{-1}$~s$^{-1}$ (see Table~\ref{table_of_values}). It is well-known that the cytoplasm can often display elevated viscosities, as high as $\eta \sim 10^{-2}$--10$^{-1}$~kg~m$^{-1}$~s$^{-1}$~\cite{LubyPhelps1999, Mogilner2018}, which would only serve to reduce $u_\textrm{ref}$ by one to two orders of magnitude (due to the scaling $u_\textrm{ref} \propto \eta^{-1}$), and result in even weaker thermal convection. Physically, for a given temperature difference (between the cell and nuclear membranes) driving the flow, a more viscous cytoplasm offers stronger resistance to motion.

Intuitively, we would of course not expect the velocity magnitudes to change drastically as the problem geometry varies. We confirm this in Fig~\ref{ecc_val_comsol_3D}, where both the maximum and minimum velocities experience only a modest change with the vertical eccentricity $e_z$ for all radius ratios $\kappa$.

\subsection{Flow structure as a function of eccentricity}

We see that the results are symmetric about $e_z=0$, which is a result of the  structure  of eqns.~\eqref{Stokes_natural_conv_ND} and \eqref{lap_T_ND}. The temperature distribution in the cell is reflected about the $z=0$ plane as $e_z$ goes from negative to positive (see Figs~\ref{e_sign}(a) and \ref{e_sign}(c)). Since the fluid flow is directly proportional to the local temperature difference, regions in the cell corresponding to the same thermal environment in the two geometries also display the same vertical velocity (Figs~\ref{e_sign}(b) and \ref{e_sign}(d)). As a result, the location of the strongest upwelling flow with respect to the nucleus changes: from being above the centre of the nucleus for $e_z < 0$, to shifting below the nucleus centre for $e_z>0$.

 \subsection{Influence of temperature difference on flow strength}

The linearity of the dimensionless flow $\ub$ with respect to $\Theta$, evident in eqn.~\eqref{Stokes_natural_conv_ND}, also lets us immediately deduce that the dimensional flow strength $\left| \ub' \right|$ varies linearly with the temperature difference $\Delta T$. This linear dependence was indeed   captured explicitly in the numerical simulations of  Ref.~\cite{Howard2019} (see their Fig~3). A clear benefit of our analysis is that, with the knowledge that heat advection can be safely neglected (i.e.~$\Pe_\textrm{t} \ll 1$), the linear relationship follows exactly mathematically.

\subsection{Summary}

In summary, assuming an axisymmetric geometry allowed us to solve for the fluid flow via a semi-analytical approach that is less computationally intensive than the finite-element simulations.  By performing two independent sets of calculations in the axisymmetric limit (finite-element  simulations and the Stokes streamfunction analysis of this section) we have confirmed the velocity magnitudes that were predicted  numerically in Sec.~\ref{numSim}. In the next section, we perform a third analysis, in which we consider the limit where the nucleus is located at the centre of the model cell (i.e.~$e_x=e_z=0$). This particular geometry enables us to obtain an exact, analytical solution for the temperature distribution and the flow field.

\section{Temperature and flow in the concentric limit}\label{concen}

\subsection{An exact solution}

When the nucleus is concentric with the cell,  one can solve the problem fully analytically in a spherical coordinate system, $\left( r, \theta, \phi \right)$~\cite{Mack1968}. Since we still have an axisymmetric setup, we can once again represent the {flow field} in terms of a streamfunction, $\Psi^\textrm{s} \left( r, \theta \right)$, where $r$ is the radial separation of any point, measured from the centre of the cell (and the nucleus), and $\theta$ is the polar angle measured from the positive $z$-axis~\cite{HappelBrenner2012}. We discuss in the main text the final results, i.e.~the analytical expressions for the temperature and the fluid velocity; all details of the solution methodology are provided in Appendix~\ref{sph_calc}. The temperature is found to be radially isotropic (i.e.~it only depends on $r$) and is given by
\begin{equation}\label{Th_conc}
    \Theta \left( r \right) = \frac{\kappa}{1-\kappa}\left( \frac{1}{r} - 1\right).
\end{equation}

As for the flow field, we derive in Appendix~\ref{sph_calc} a very simple representation for the streamfunction $\Psi^\textrm{s} \left( r, \theta \right)$ as
\begin{equation}\label{psi_simplified}
    \Psi^\textrm{s}(r,\theta) = f_1(r) \sin^2 \theta,
\end{equation}
where  
\begin{equation}\label{f1r}
    f_1(r) = \frac{r^3}{8}\frac{\kappa}{1-\kappa} + c_1 r + \frac{c_2}{r} + c_3 r^2 + c_4 r^4,
\end{equation}
with the constants, $c_i$, being functions solely of the radius ratio $\kappa$ given by
\begin{align}\label{c_i_f1r}
    c_1 &= \frac{\kappa^2(3\kappa^2 + 4\kappa + 3)}{8(1-\kappa)(4\kappa^2 + 7\kappa + 4)}, \\ \nonumber
    c_2 &= -\frac{\kappa^4}{8(1-\kappa)(4\kappa^2 + 7\kappa + 4)}, \\ \nonumber
    c_3 &= -\frac{\kappa(\kappa+1)(\kappa^2 + 3\kappa + 1)}{4(1-\kappa)(4\kappa^2 + 7\kappa + 4)}, \\ \nonumber
    c_4 &= -\frac{\kappa(\kappa+1)}{4(1-\kappa)(4\kappa^2 + 7\kappa + 4)}.
\end{align}
Once the streamfunction is known, the fluid velocity $\ub$ may be obtained explicitly as
\begin{align}\label{ur_umu_1}
    \ub = \left( u_r, u_{\theta} \right) = \left( -\frac{1}{r^2 \sin \theta} \frac{\partial \Psi^\textrm{s}}{\partial \theta}, \frac{1}{r \sin \theta} \frac{\partial \Psi^\textrm{s}}{\partial r} \right),
\end{align}
where $u_r$ is the radial component and $u_{\theta}$ the tangential component of the fluid velocity. Substituting eqn.~\eqref{psi_simplified} in \eqref{ur_umu_1} yields,
\begin{align}\label{ur_umu_2}
    u_{r}\left( r, \theta \right) &= -\frac{2 f_1\left( r \right)}{r^2} \cos \theta, \\ \nonumber
    u_{\theta}\left( r, \theta \right) &= \frac{1}{r} \frac{df_1}{dr} \sin \theta.
\end{align}
We thus have a closed-form, exact solution to the temperature-gradient-driven natural convection problem in the concentric case, for arbitrary radius ratios $\kappa = R_\textrm{nuc}/R_\textrm{c}$. A simple transformation from spherical to Cartesian coordinates  then yields the vertical velocity $u_z$ and the horizontal velocity $u_x$ (see eqn.~\eqref{uz_ux} in Appendix~\ref{sph_calc}). In Fig~\ref{Sph_vs_COMSOL}, we plot these velocity components (left column) and compare them with the corresponding COMSOL simulation results (right column). We can see an excellent agreement between the exact solution and our computations; quantitatively, the average relative error between the two solutions is less than 2\%, which provides another validation of the simulations of Sec.~\ref{numSim} and confirms that the velocity predictions therein are accurate.

\begin{figure}[t]
\begin{center}
\includegraphics[width=12cm]{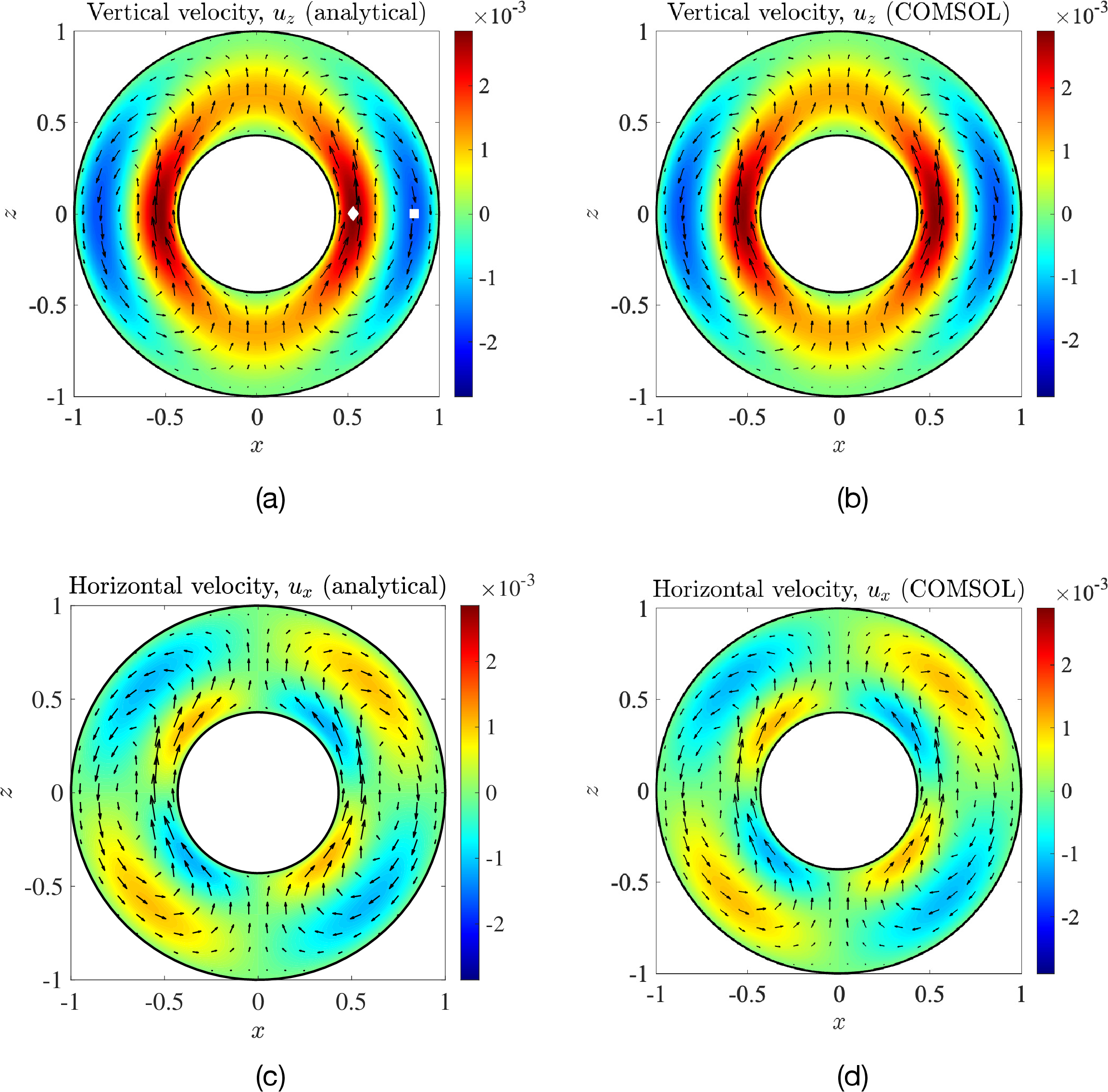}
\caption{Dimensionless vertical component of the flow field at the $y=0$ mid-plane (panels (a) and (b)) and horizontal component (panels (c) and (d)), calculated analytically using eqn.~\eqref{uz_ux} (left column) and via COMSOL simulations (right column): geometrical parameters are $\kappa = 0.43, e_x = 0, e_z = 0$. The symbols in panel (a) are locations of the maximum and minimum velocities, whose nearest distance from the cell membrane (i.e.~from the point $x=1$, $y=0$, $z=0$) is plotted in Fig~\ref{max_min_vel_details}(c).
The arrows indicate the in-plane velocity.}

\label{Sph_vs_COMSOL}
\end{center}
\end{figure}

\subsection{Structure of flow in concentric geometry}

\begin{figure}[t]
\begin{center}

\includegraphics[width=\linewidth]{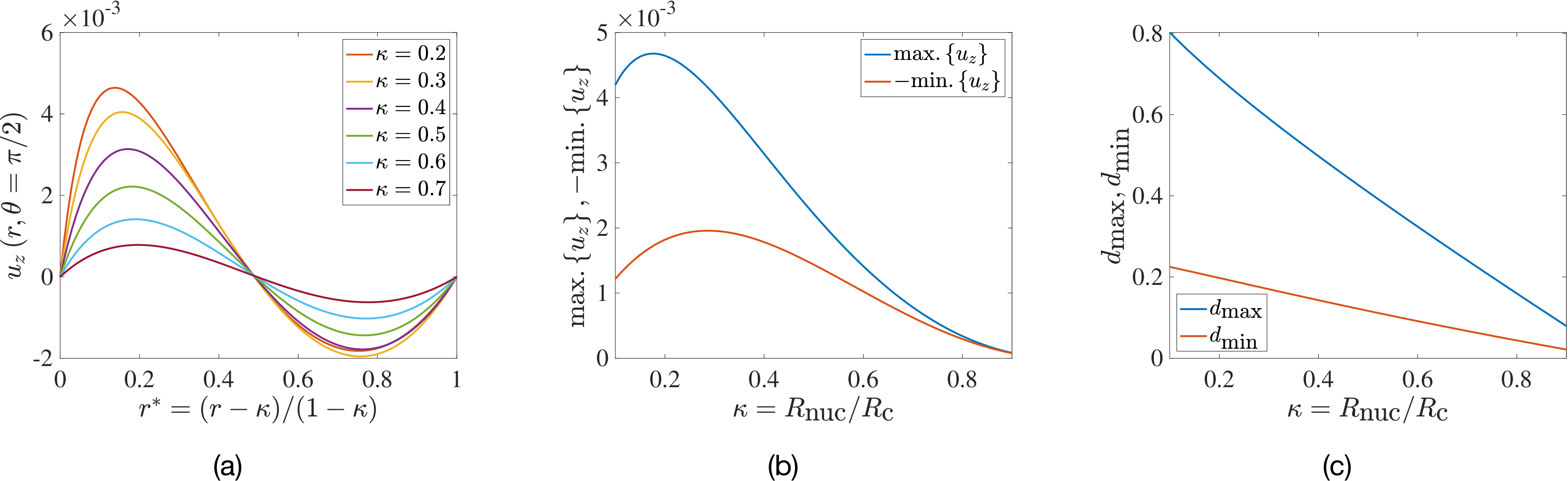}

\caption{Upwelling and downwelling {convective flows}. (a) Dimensionless vertical fluid velocity at the equatorial plane, plotted as a function of normalised radial distance from the nucleus, $r^*$. (b) Variation with the radius ratio, $\kappa$, of the magnitudes of the strongest upwelling $\left( \textrm{max}. \left\{ u_z \right\} \right)$ and downwelling $\left( -\textrm{min}. \left\{ u_z \right\} \right)$ flows. (c) Radial separations $d_\textrm{max}$ and $d_\textrm{min}$, from the cell membrane of the regions of maximum  and minimum velocity, respectively.}
\label{max_min_vel_details}
\end{center}
\end{figure}

Because of symmetry about the $z$-axis, the fluid from the right-half of the cell cannot flow into the left-half. This, combined with mass conservation,  means that the  {net} flow-rate through the annulus, $Q_\textrm{ann} \sim \int_{\kappa}^{1} {u_z(r,\theta=\pi/2) r \,dr}$, has to be exactly zero. Hence, any upwelling flow perpendicular to the equatorial plane ($z=0$, alternatively $\theta = \pi/2$) must be cancelled out, in an averaged sense, by a downwelling flow. This can {clearly be seen} in Fig~\ref{Sph_vs_COMSOL}(a). The (vertical) velocity at the equatorial plane is given by
\begin{equation}\label{uz_eq}
u_z \left( r, \theta = \pi/2 \right) = -u_{\theta} \left( r, \theta = \pi/2 \right) = -\frac{1}{r}\frac{df_1}{dr},
\end{equation}
which we plot in Fig~\ref{max_min_vel_details}(a) as a function of a normalised separation from the nucleus surface, $r^* \equiv{(r - \kappa)}/{(1 - \kappa)}$; thus, $r^*=0$ corresponds to the nucleus surface, while $r^*=1$ corresponds to the cell membrane. 

For each value of $\kappa$, the maximum and minimum velocities in Fig~\ref{max_min_vel_details}(a) correspond to the strongest upwelling and downwelling flows, respectively, in the entire cell. We further plot in Fig~\ref{max_min_vel_details}(b)  the magnitude of these {strongest} vertical flows as a function of the radius ratio $\kappa$, while their position, as measured radially from the cell membrane, {is displayed} in Fig~\ref{max_min_vel_details}(c). We observe a clear asymmetry in the flow as seen in Figs~\ref{max_min_vel_details}(a) and \ref{max_min_vel_details}(b); in terms of magnitude, the strongest upwelling flow is greater than the strongest downwelling flow.  
This asymmetry is a consequence of spherical geometry: the differential area through which the fluid flows, $dA(r) = 2 \pi r dr$, increases with the radius $r$, which necessitates that the fluid near the nucleus must rise faster than the fluid near the cell membrane settles, in order to maintain zero net-flux in the annulus. This asymmetry {thus vanishes} when $\kappa \to 1$.

 \subsection{Influence of a more general cell geometry}

The results in eqns.~\eqref{psi_simplified} to \eqref{ur_umu_2} give us an exact representation of the flow inside the cell, in this simple geometrical case. How relevant are these predictions for arbitrary locations of the nucleus? Further computations shown in Fig~\ref{infl_nonaxisymm} (Appendix~\ref{infl_geom}) demonstrate that the  strength  of the flow depends only weakly on the position of the  nucleus. Therefore, the exact  analysis of this section captures all essential aspects of this hydrodynamics problem.

 \subsection{Summary}

We have now validated our numerical simulations using two independent analyses, which gives us confidence in our numerical results from Sec.~\ref{numSim}. We thus conclude that the flow strengths that we report are a more accurate estimate of intracellular buoyancy-driven flows than the results of Ref.~\cite{Howard2019}.

\section{Solute transport by combined convection and diffusion}\label{solTrans}

Having solved for the fluid flow numerically, semi-analytically and fully analytically, we have concluded that temperature differences between a cell's nucleus and cell membrane lead to much weaker flows than envisioned previously. Importantly, our analysis shows that quantitative predictions on transport cannot be made solely using scaling arguments, but that they instead require solving for the details of the flow field. Indeed, the standard reference velocity $u_\textrm{ref} = \rho_0 \beta g \Delta T R^2_c/\eta${, obtained by scaling arguments,} predicts typical velocities that are  {around} 0.5 $\mu$m~s$^{-1}$ (for $\Delta T = 1\;$K), whereas  {full} calculations yield values of $\sim O(10^{-3})$~$\mu$m~s$^{-1}$, which are two orders of magnitude smaller.

We now investigate the ability of these flows to transport solute inside the cell, e.g.~proteins and other macromolecules~\cite{Mogre2020}.

\subsection{Solute transport with both diffusion and advection}

We model the solute as a passive scalar that diffuses and is advected by the flow calculated in Sec.~\ref{numSim}. At steady state, the dimensionless concentration of the scalar, $C$, is governed by the advection-diffusion equation
\begin{equation}\label{adv_diff}
    \ub \cdot \nabla C = \frac{1}{\Pe_\textrm{s}} \nabla^2 C,
\end{equation}
where
$\Pe_\textrm{s} = u_\textrm{ref}R_\textrm{c}/D$ is the P\'eclet number of the solute (the ratio of its advective to its diffusive transport), where $D$ is the solute's molecular diffusivity. For the sake of simplicity, we  assume that the scalar is produced at the nucleus and absorbed at the cell membrane, so  we have the boundary conditions
\begin{align}\label{scalar_BCs}
    C_\textrm{nuc} &= 1, \\ \nonumber
    C_\textrm{mem} &= 0.
\end{align}
Typical molecular diffusivities of cellular matter can be as low as 0.01~$\mu$m$^2$~s$^{-1}$~\cite{LubyPhelps1999, Goldstein2015}, which corresponds to 
$\Pe_\textrm{s} \approx 500$. Based on this preliminary scaling approach, one may then expect the cell-scale natural convection described previously to significantly affect solute transport. 

\subsection{Advection-enhanced transport: the local P\'eclet number}

\begin{figure}[t]

\includegraphics[width=\linewidth]{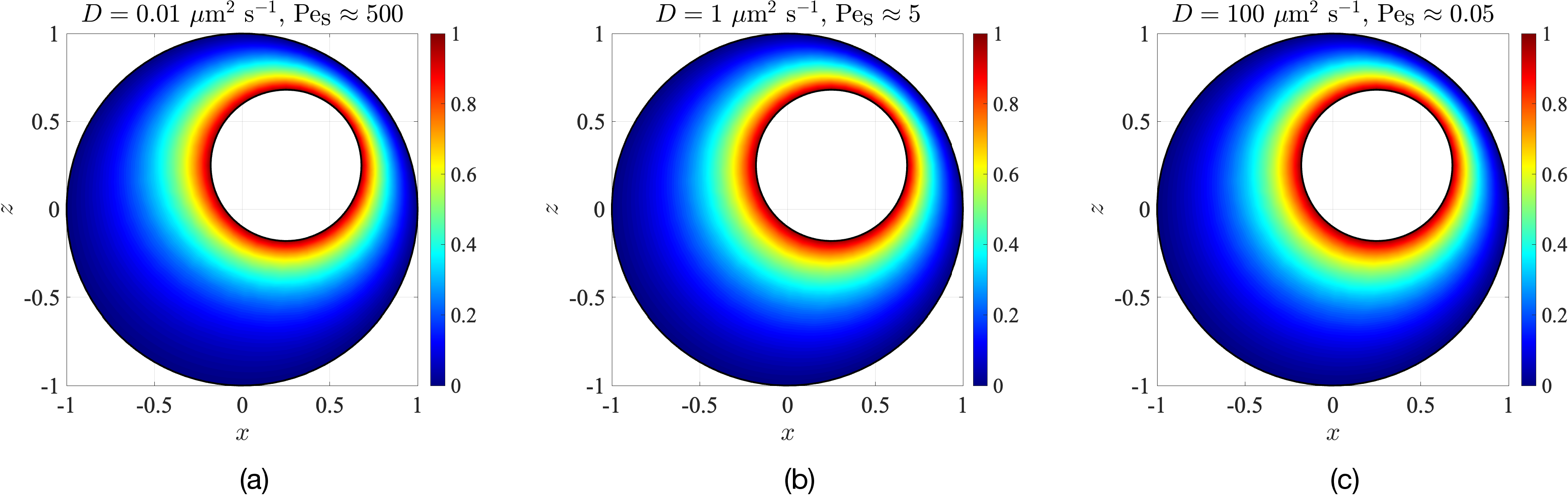}

\caption{Influence of the temperature-gradient-driven convection flow on the distribution of a passive scalar with three different diffusivities, (a)~$D=0.01$~$\mu$m$^2$~s$^{-1}$, (b)~$D=1$~$\mu$m$^2$~s$^{-1}$ and (c)~$D=100$~$\mu$m$^2$~s$^{-1}$.}
\label{advection_infl}
\end{figure}

To investigate the extent to which this is true,
we solve eqns.~\eqref{adv_diff} and \eqref{scalar_BCs} for the configuration and parameters of Sec.~\ref{numSim} $\left( \kappa = 0.43, e_x=e_z=0.25, \Delta T = 1\;\textrm{K} \right)$ using finite-element COMSOL simulations. The resulting solute distributions on the $y=0$ mid-plane are shown in Fig~\ref{advection_infl}. Surprisingly, the effect of natural convection is so weak that even for $\Pe_\textrm{s}$ as large as 500 (Fig~\ref{advection_infl}(a)), the solute distribution is almost the same as that when the solute transport is purely diffusive, i.e.~when $\Pe_\textrm{s} \ll 1$ (Fig~\ref{advection_infl}(c)). One can also visualise the time evolution of the concentration, starting from a uniform initial condition $C\left( t' = 0 \right) \equiv 0$ and then suddenly changing the concentration at the nucleus surface to $C_\textrm{nuc}=1$ at $t'=0.1$~s; this is done in Appendix~\ref{c_unst_evol}. As shown in Fig~\ref{c_vs_t} there, the concentration distribution becomes nearly independent of $\Pe_\textrm{s}$ when the time is normalised by the diffusive time-scale $t'_\textrm{d} = R_\textrm{c}^2/D$, which is a classic signature of isotropic diffusion. This further confirms that the transport mechanism inside the cell is mostly diffusive.

For given size and material properties of the cell, the definition of $\Pe_\textrm{s}$ (see Table \ref{table_of_values}) implies that the main factors influencing solute transfer are the temperature difference~$\Delta T$ and the solute diffusivity~$D$. To quantify their effects, we may   define a local P\'eclet number {for the solute}, Pe$^\ell=2 \left| \ub' \right| R_\textrm{c}/D$, with maximum value Pe$^\ell_\textrm{max}=2 u'_\textrm{max} R_\textrm{c}/D$. Recall here that $\ub'$ is the dimensional fluid velocity, which is a function of position within the cytoplasm, while $u'_\textrm{max} \equiv \textrm{max}. \{ | \ub' | \}$ is the global maximum magnitude of the fluid velocity. 

\begin{figure}[t]
\begin{center}

\includegraphics[width=\linewidth]{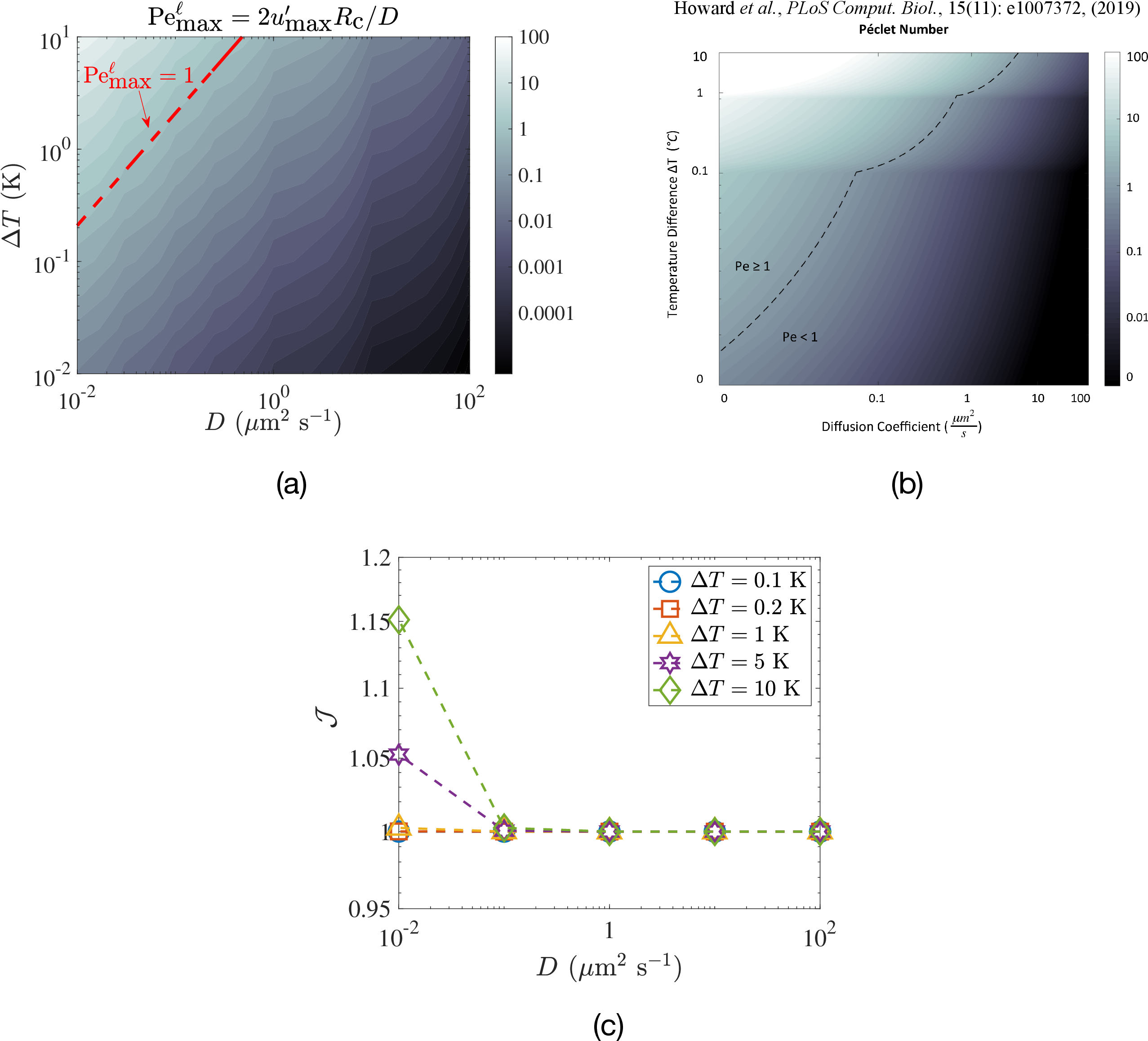}

\caption{Quantifying  transport arising from {convective flows}. (a) Maximum local P\'eclet number, Pe$^\ell_\textrm{max}$, as a function of temperature difference $\Delta T$ and solute diffusivity $D$, obtained through non-linear simulations in COMSOL. (b) Same maximum local P\'eclet number as obtained numerically in Ref.~\cite{Howard2019}. (c) Normalised solute flux $\mathcal{J}$ as a function of $D$ for different prescribed temperature differences. In all cases, the geometry corresponds to that in Fig~\ref{advection_infl}.}
\label{Pe_solute}
\end{center}
\end{figure}

In order to understand the weak impact of convective flows on transport, we plot in in Fig~\ref{Pe_solute}(a)  the iso-values of the maximum local  P\'eclet number, Pe$^\ell_\textrm{max}$,  for different sets of temperature difference ($\Delta T$) and diffusivity ($D$). We also contrast  our computational results with the same quantity obtained in Ref.~\cite{Howard2019} in their simulations and reproduce their results in Fig~\ref{Pe_solute}(b). Once again, it is notable that our results have  local P\'eclet numbers about one order of magnitude smaller than those in  Ref.~\cite{Howard2019}.
 
 \subsection{Absorption flux across nuclear membrane}

While {the maximum local P\'eclet number} $\Pe^\ell_\textrm{max}$ captures a local estimate of mass transfer enhancement by cytoplasmic convection,  a fairer quantification of this enhancement at the whole-cell level involves calculating the total absorption flux  across the nuclear membrane, and comparing it to the limit when the solute emitted by the nucleus {instead} undergoes pure diffusion. If $S$ denotes the surface of the nucleus, {then} the dimensionless diffusive flux through the surface is given by the integral $\int_{S} { -\mathbf{n} \cdot \nabla C \,dS }$, while the dimensionless advective flux, $\int_{S} { \mathbf{n} \cdot \ub C  \,dS }$, vanishes due to impermeability of the nuclear surface (see eqn.~\eqref{u_BCs}).  {The influence of bulk solute advection on the mass transfer rate across the nucleus is quantified by defining a normalised flux,}
\begin{equation}\label{Sherwood_like}
    \mathcal{J}=\frac{\int_{S} \mathbf{n} \cdot \nabla C \,dS}{\left. \left( \int_{S} \mathbf{n} \cdot \nabla C \,dS \right) \right|_{\Pe_\textrm{s} \equiv 0}},
\end{equation}
{where the denominator is the solute flux through the nucleus when the solute transport is purely diffusive (i.e.~when $\Pe_\textrm{s} \equiv 0$ in eqn.~\eqref{adv_diff}). In the mass transfer literature, the quantity $\mathcal{J}$ is usually referred to as the Sherwood number~\cite{Leal2007} and is expected to be a function of the cell geometry and the solutal P\'eclet number, $\Pe_\textrm{s}$.}

 \subsection{Convective flows are too small to provide significant enhancement  of absorption flux}

If bulk advection of solute were to significantly impact the solute  transfer from the nucleus to the cell membrane, then  we would expect  $\mathcal{J}$ to be greater than unity. We plot in Fig~\ref{Pe_solute}(c) the dependence of this normalised flux on the diffusivity of the solute ($D$) for various temperature differences ($\Delta T$). The solute diffusivity $D$ spans four orders of magnitude, covering cellular material ranging from large vesicles ($D \sim 10^{-2}$~$\mu$m$^{2}$~s$^{-1}$~\cite{Felder1994, Burke1997}), to proteins synthesised in the ribosomes ($D \sim 1$~$\mu$m$^{2}$~s$^{-1}$ to $10$~$\mu$m$^{2}$~s$^{-1}$~\cite{LubyPhelps1999, GuraSadovsky2017}), to metabolites such as ATP, ADP and calcium ions ($D \sim 100$~$\mu$m$^{2}$~s$^{-1}$~\cite{Donahue1987, Vendelin2000}).

For the lowest $\Delta T$ value considered, the convective flows do not provide any advantage in driving the solute. Even for $\Delta T$ as large as 10~K (a large value not supported by experimental data) and solute diffusivities as low as 0.01 $\mu$m$^2$~s$^{-1}$ (relevant for the largest vesicles in the cell), the solute removal from the nucleus is only amplified by about 15\% compared with the case of pure diffusion.

As an example, consider the protein insulin, whose diffusivity in living cells is around 1~$\mu$m$^{2}$~s$^{-1}$~\cite{LubyPhelps1999}. The result in Fig~\ref{Pe_solute}(c) then tells us that $\mathcal{J} \sim 1$  for any range of experimentally observed intracellular temperature differences. Most cellular proteins and macromolecules have diffusivities that are too large (ranging from 1~$\mu$m$^{2}$~s$^{-1}$ to 10~$\mu$m$^{2}$~s$^{-1}$) to be overcome by natural convection. Hence, their motion through the cell likely  remains diffusive. Vesicles in cells can have very low diffusivities, around $10^{-2}$~$\mu$m$^{2}$~s$^{-1}$~\cite{Felder1994, Burke1997} but, despite this, for the commonly measured temperature differences within cells, their transport enhancement due to convective flows is expected to remain, at best, very modest, and most likely, completely negligible.

\section{Discussion}\label{conclusion}

\subsection{Intracellular convective flows are smaller than previously predicted}\label{recap_flow}

In this work, we re-considered the problem of intracellular natural convection potentially induced by a temperature difference $\Delta T$ between the cell's nucleus and membrane. Specifically, we used numerical simulations to show that intracellular flows are expected to be on the order of $\sim 10^{-3}$~$\mu$m~s$^{-1}$ in magnitude for a nucleus-to-membrane temperature difference of 1~K. In the process, we discovered an important disparity from previous results that had predicted ten-fold stronger flows for the same geometry and temperature differential~\cite{Howard2019}. 

We explored further this discrepancy, and confirmed it, by performing independent calculations of the flow strength in an axisymmetric geometry (Sec.~\ref{axisymm}) and in a concentric geometry (Sec.~\ref{concen}). Importantly, in the latter case, we provided an exact expression for the flow inside the cell. All our analyses converged to a similar value for the flow strength, which was indeed $\sim 10$ times weaker than that computed in Ref.~\cite{Howard2019}.  These results highlight the importance of detailed flow calculations, since simple  scaling analysis alone suggests fluid velocities almost two orders of magnitude larger ($u_\textrm{ref} \sim 0.5$~$\mu$m~s$^{-1}$), which would lead one to believe, incorrectly, that these convective flows are very strong.

\subsection{Intracellular convection does not lead to significant increase in chemical transport beyond diffusion}

The transport of materials within a cell is essential to  its normal function, from  vesicle transport for structural upkeep~\cite{Fishman2003}, signal transmission via proteins~\cite{GuraSadovsky2017} and organelle transport during cell division~\cite{Mishra2014, Knoblach2016} to inter-organelle interactions~\cite{Cohen2018} and the maintenance of nutrient/metabolite gradients~\cite{Niethammer2008}. The movement of cellular matter is accomplished by a variety of mechanisms, including molecular diffusion, active transport by motor proteins and advection due to cytoplasmic flows~\cite{Mogre2020}. A second aim of our work was to investigate whether temperature-gradient-driven flows   could contribute usefully to advective material transport inside the cell. This would be particularly relevant for the motion of cell constituents with low diffusivities (large organelles and vesicles), which necessarily require alternative mechanisms to traverse the cell. Towards this, we numerically simulated the mass transfer of a chemical species by cytoplasmic convection, and showed that convection does not contribute significantly to the transport of material within a cell beyond what is achieved by pure diffusion. We quantified the advection-induced enhancement in averaged mass-flux of a chemical species released from the nucleus and absorbed at the cell membrane, as a function of the prescribed temperature difference and the diffusivity of the species. Only in situations that are  biologically unrealistic (temperature differences of 10~K and very small diffusivities of 0.01~$\mu$m$^2$~s$^{-1}$, associated with cellular vesicles) could the averaged mass-flux see a modest increase above  that obtained  with purely diffusive mass transfer.

Thus, while a cell's thermal environment is important for its metabolism~\cite{Takeuchi2009, Wang2020} and survival~\cite{Richter2010}, the flows generated by temperature gradients, in most realistic scenarios, have negligible effect in improving intracellular mass transfer. It seems therefore that these flows do not noticeably impact important cell processes like protein delivery, signal propagation, and organelle and metabolite transport.

Once again, this conclusion becomes apparent only after a detailed calculation of solute transport in the bulk, as done in Sec.~\ref{solTrans}. A simple scaling analysis yields characteristic  {solutal}  P\'eclet numbers $\Pe_\textrm{s} = u_\textrm{ref}R_\textrm{c}/D \approx 500$ that are much larger than unity, which, in the absence of  detailed numerical calculations, would erroneously suggest a significant natural-convection-induced advective contribution to species transport.

 \subsection{A broader range of modelling assumptions could be explored in future work}

One of the reasons for the aforementioned flows being so weak is that we have modelled the nucleus surface and the cell membrane as rigid and non-slipping, i.e.~the fluid's tangential velocity vanishes at both these surfaces. As a result, there are significant viscous stresses in the domain that resist fluid motion. A different model could allow   tangential motion along the cell surface due to slip or mobility of the membrane.  

Our model simplifies the cytoplasm and assumes {that} it is effectively a homogeneous continuum with Newtonian properties. It is obviously known to be more complex,  consisting of a polymeric and dynamic cytoskeleton embedded in a viscoelastic, gel-like fluid that flows past freely suspended cell organelles~\cite{Mogilner2018}.

Our analysis also considers a simplified description of thermal diffusion inside the cell. Inherent in eqn.~\eqref{advDiff_T} is the assumption that the thermal diffusivity of the cytoplasm is isotropic, i.e.~that heat diffuses at a constant rate along all directions. Since the cytoplasm is  heterogeneous, the thermal diffusivity within it is expected to vary spatially~\cite{Song2021}. 

\subsection{Our model could be adapted to quantify artificially induced convection}

Our calculations show that cytoplasmic flows due to naturally occurring temperature gradients within a cell are not very effective in driving cellular material transport. However, artificially induced thermal convection has been recently proposed as a means to accelerate cell assembly for biomedical assays~\cite{Zhang2023}.  Light-absorbing particles are arrested in optical traps, are heated in the process and thus induce thermal flows that drive cell accumulation around the particles.  One could envision the generation of such flows within a cell as well, for example, by laser-assisted heating of metallic nano-particles~\cite{Govorov2007, Donner2011, Roxworthy2014}.  Our theoretical analysis could thus be adapted to quantify such artificially induced cell-scale convection, to inform future hydrodynamics-based strategies for intracellular object manipulation.

\section*{Appendix}
\appendix

\section{Fluid velocity along the $y'$ direction}\label{y_velocity}

\begin{figure}[t]
\begin{center}

\includegraphics[width=\linewidth]{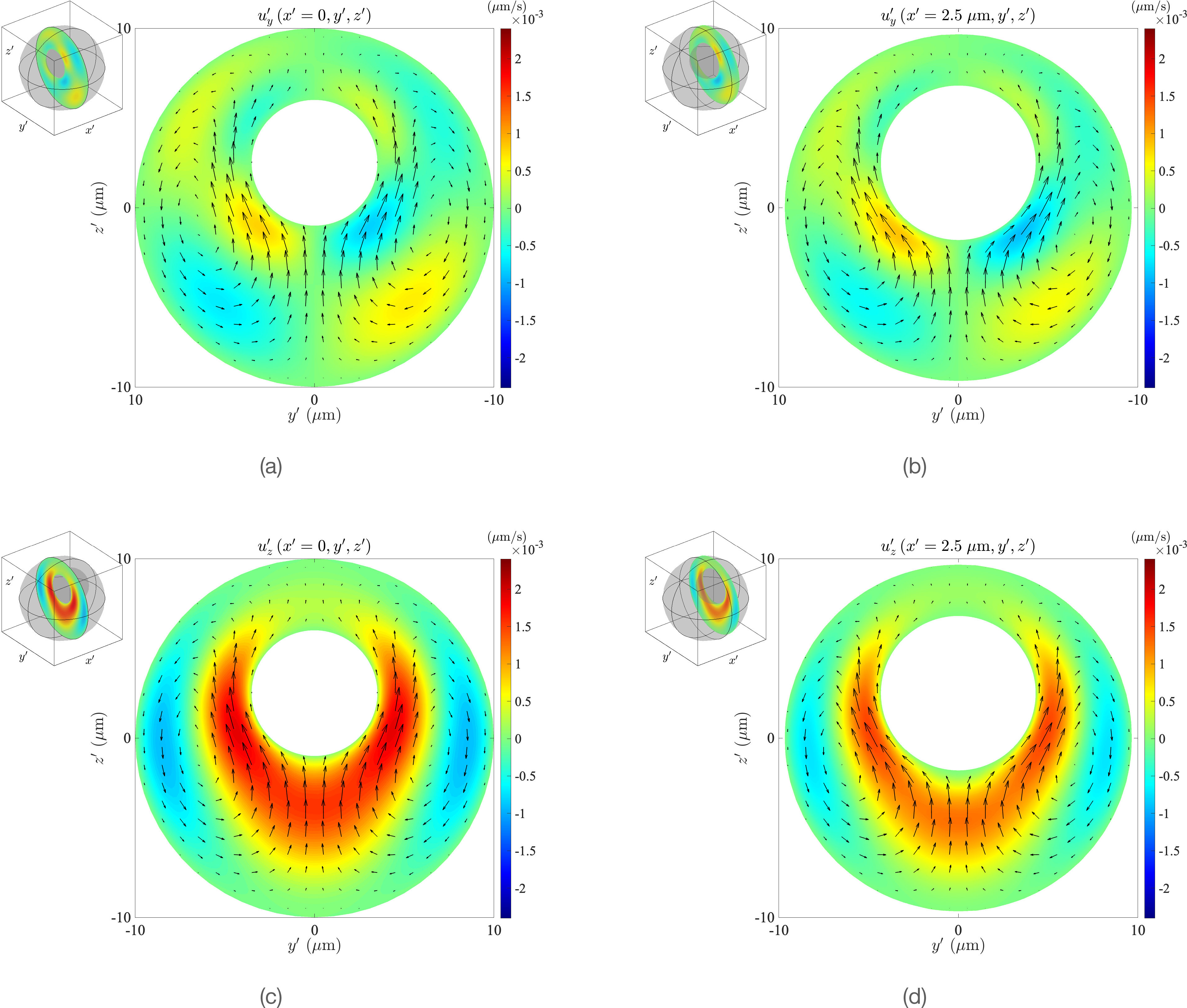}

\caption{The $y'$  component  of the fluid velocity (panels (a) and (b)) and the $z'$ component  (panels (c) and (d))  at vertical sections through the cell centre (panels (a) and (c)) and the nucleus centre (panels (b) and (d)). The arrows denote in-plane velocity vectors. Note that the $y'$-axis is inverted in panels (a) and (b), for ease of visual comparison with the contours seen in the insets. The cell geometry is the same as in Fig~\ref{vel_contours_AD_vs_oD} $\left( \kappa = 0.43,\;e_x=e_z=0.25 \right)$.}

\label{y_vel_slices_x}

\end{center}
\end{figure}

\begin{figure}[t]
\begin{center}

\includegraphics[width=12cm]{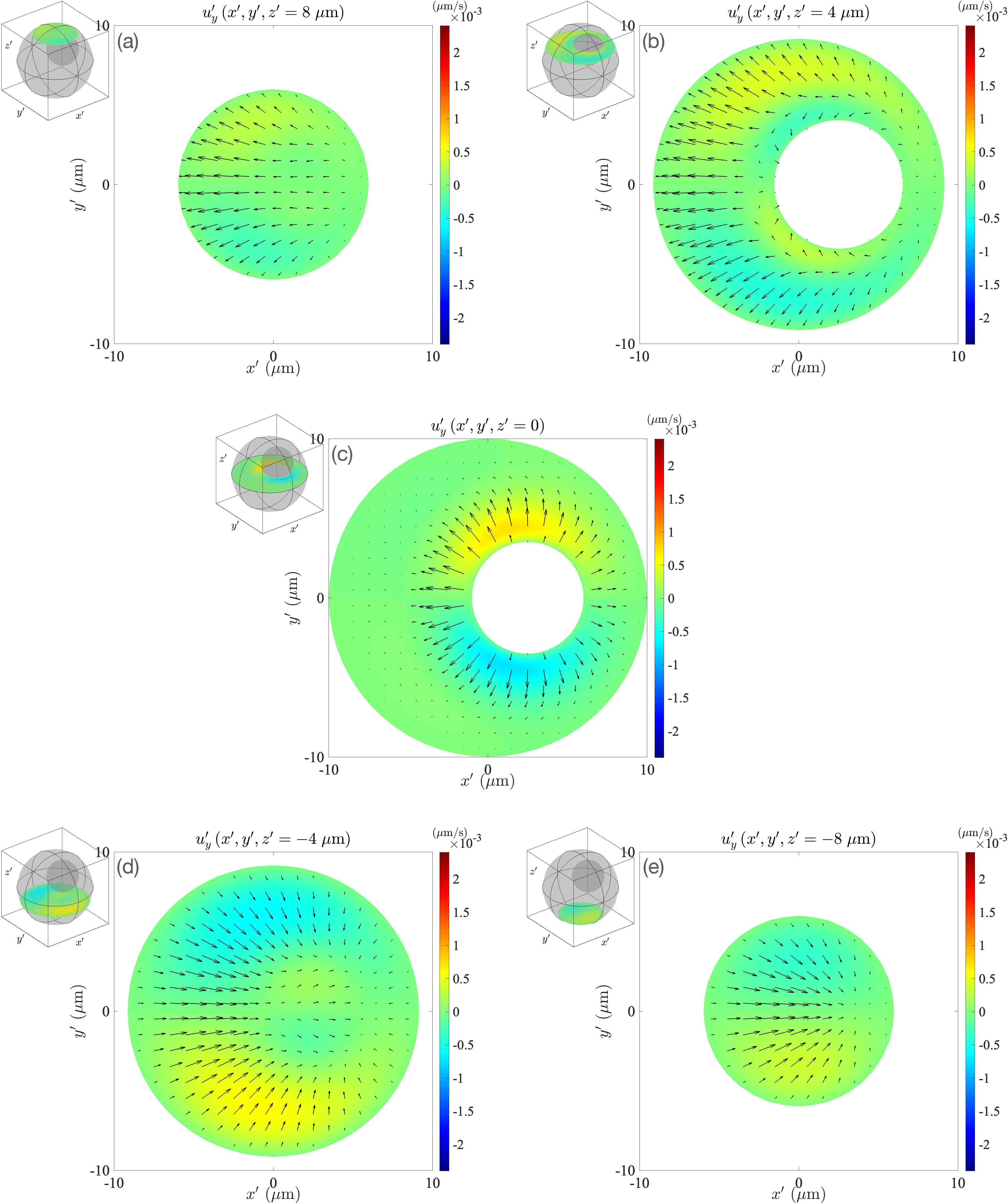}

\caption{The $y'$ component of the fluid velocity at different horizontal sections through the cell, as shown in the inset in each figure. The arrows denote in-plane velocity vectors. The cell geometry is the same as in Fig~\ref{vel_contours_AD_vs_oD} $\left( \kappa = 0.43,\;e_x=e_z=0.25 \right)$.}

\label{y_vel_slices_z}

\end{center}
\end{figure}

In the main text, we discussed the nature of fluid flow in the vertical plane through the centres of both the cell and the nucleus (the $x'$-$z'$ plane). Here we discuss  the importance of the other component of the fluid velocity, i.e.~the velocity $u'_y$, in comparison with $u'_x$ and $u'_z$.  In Fig~\ref{y_vel_slices_x} we show this  horizontal velocity $u'_y$ on two vertical planes $x'=\text{const.}$ -- through the centre of the cell and through the centre of the nucleus -- along with the vertical velocity $u'_z$. Similarly, Fig~\ref{y_vel_slices_z} shows $u'_y$ on five different horizontal planes through the cell, at different heights $z'$. The cell geometry and the physical parameters in both these figures are the same as in Fig~\ref{vel_contours_AD_vs_oD}. Note that since only the full three-dimensional flow is incompressible, the in-plane velocity in Figs~\ref{y_vel_slices_x} and \ref{y_vel_slices_z} is not incompressible; this is most apparent in Fig~\ref{y_vel_slices_z}, where we can see local two-dimensional  sources and sinks of flow. Clearly, the typical magnitude of $u'_z$ is larger than $u'_y$ in both Figs~\ref{y_vel_slices_x} and~\ref{y_vel_slices_z} (see also Fig~\ref{vel_contours_AD_vs_oD}). This means that the strongest flows in the cell are vertical (along or opposite to gravity) and justifies our focus on $u'_z$  in  the main text.

\clearpage

\clearpage

\section{Influence of nucleus position on fluid flow: COMSOL simulations}\label{infl_geom}

\begin{figure}[t]
\begin{center}

\includegraphics[width=\linewidth]{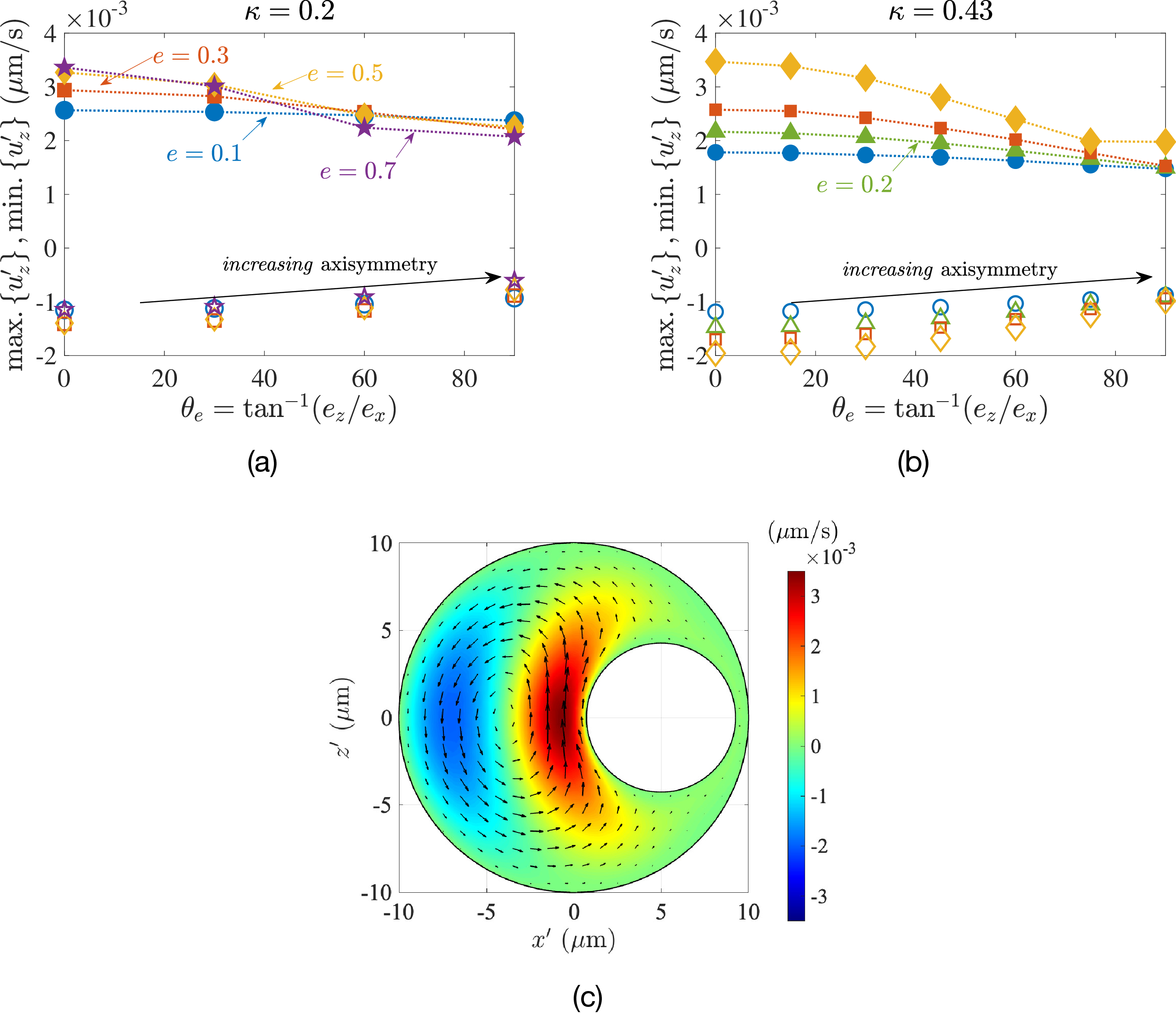}

\caption{Impact of the extent of axisymmetry (quantified by $\theta_e$, shown here in degrees) on the dimensional maximum upwelling (filled symbols) and minimum downwelling (open symbols) flows inside the cell, for (a) $\kappa=0.2$, (b) $\kappa=0.43$. The marker types and colours denote different eccentricity values, whose legend is the same across both panels. The temperature difference between the nucleus and the cell membrane in each case is $\Delta T=1$\;K. Panel (c) further shows the vertical velocity in the mid-plane for $\theta_e=0$, $e=0.5$ (i.e.~$e_x=0.5, e_z=0$) and $\kappa = 0.43$. The arrows indicate the in-plane velocity.}
\label{infl_nonaxisymm}
\end{center}
\end{figure}

In this section, we use numerical simulations to briefly summarise the dependence of the flow on the position of the nucleus inside the cell. The angle $\theta_e = \tan^{-1} (e_z/e_x)$ shown in Fig~\ref{prob_schem} quantifies how axisymmetric the system is, with $\theta_e = \pi/2$~rad~($=90\degree$) representing the axisymmetric case. Results plotted in Fig~\ref{infl_nonaxisymm} show that, for all values of the eccentricity $e$, the flow strength increases monotonically as the extent of axisymmetry decreases. Intuitively, for $\theta_e = 0$, the majority of the fluid is least confined in the direction normal to gravity. Hence, for any non-zero eccentricity/offset, the fluid experiences the least viscous resistance when $\theta_e = 0$. As a result, it attains the largest velocities for a prescribed driving temperature difference. This can be seen in Fig~\ref{infl_nonaxisymm}(c), where the fluid to the left of the nucleus is heated and attains large upwelling speeds.

However, the increase in the dimensional flow speed $\left| \ub' \right|$ as $\theta_e$ decreases is only modest, and  the maximum velocities inside the cell still remain $\sim 10^{-3}$~$\mu$m~s$^{-1}$. Since the order of magnitude of the flow inside the cell does not change over the entire range of parameters, the predictions from Secs.~\ref{axisymm} and \ref{concen} provide a quantitative estimate of the convective flows occurring in  more complex configurations.

\section{Time-evolution of concentration inside the cell}\label{c_unst_evol}

\begin{figure}[h]
\begin{center}

\includegraphics[width=11cm]{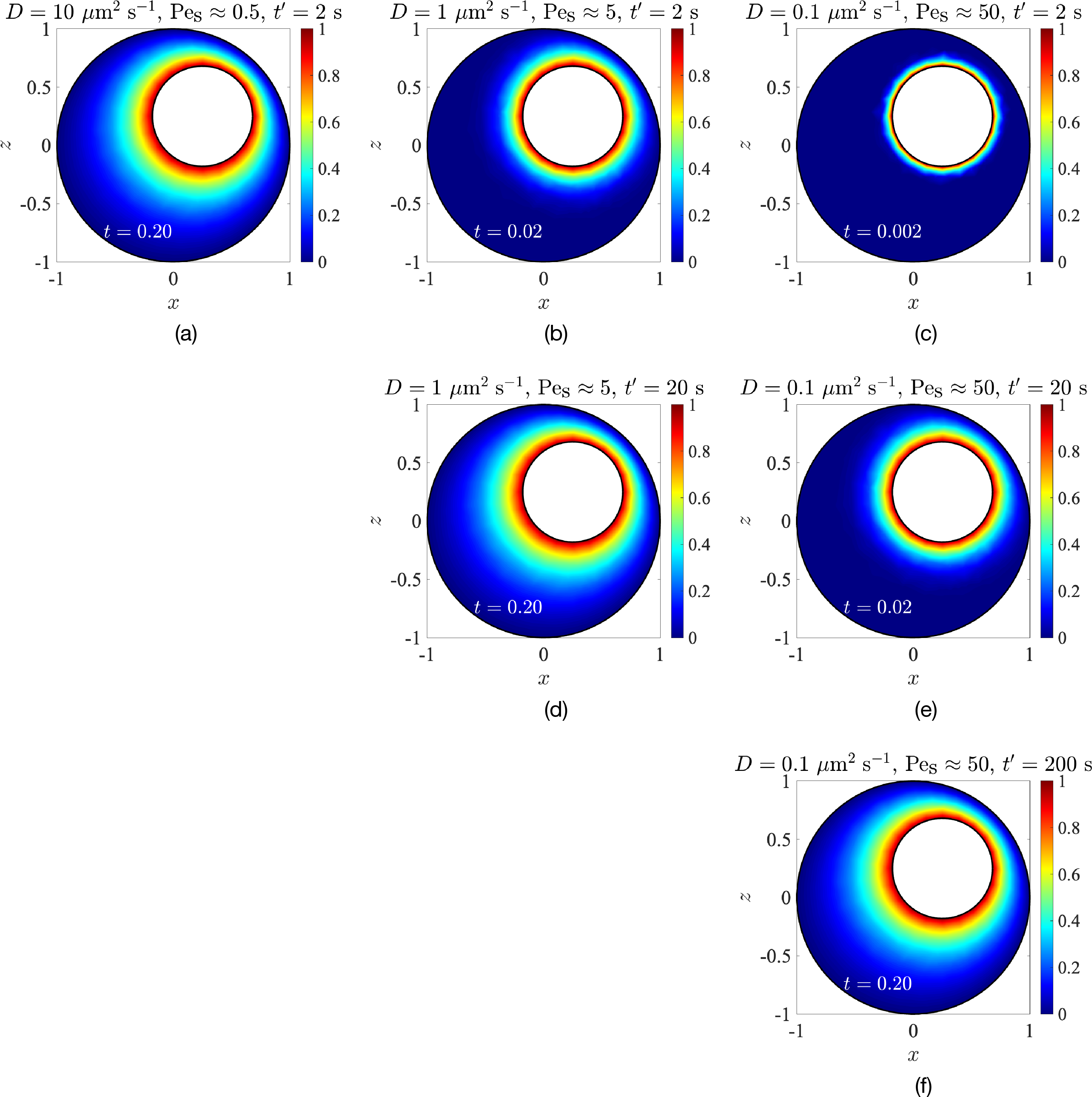}
    
 \caption { Influence of the temperature-gradient-driven convection flow on the unsteady intracellular distribution of a passive scalar with three different diffusivities, $D=10$~$\mu$m$^2$~s$^{-1}$ (left column), $D=1$~$\mu$m$^2$~s$^{-1}$ (middle column) and $D=0.1$~$\mu$m$^2$~s$^{-1}$ (right column). The top row corresponds to dimensional time $t'=2$~s, the middle row, $t'=20$~s and the bottom row, $t'=200$~s. Note the equivalence between panels (a), (d) and (f), and also between (b) and (e), which correspond to the same dimensionless time $t$.}
    
\label{c_vs_t}
    
\end{center}
\end{figure}

While there is very little difference in the steady-state concentration profiles in Fig~\ref{advection_infl}, the time-evolution of concentration shows more discernible changes between different values of $\Pe_\textrm{s}$. In Fig~\ref{c_vs_t}, we show results from COMSOL simulations, where the initial concentration is uniform everywhere inside the cell, set as $C \left( \textbf{x}', t'=0\right) =0$, and then, at dimensional time $t'=0.1$~s, the concentration at the nucleus surface is changed to $C_\textrm{nuc}=1$. These simulations were performed for three values of solute diffusivity: $D = 10$~$\mu$m$^2$~s$^{-1}$ (Pe$_\textrm{s} \approx 0.5$), $D = 1$~$\mu$m$^2$~s$^{-1}$ (Pe$_\textrm{s} \approx 5$) and $D = 0.1$~$\mu$m$^2$~s$^{-1}$ (Pe$_\textrm{s} \approx 50$). In each case, the solute concentration evolves to a steady state over the diffusive time-scale $t'_\textrm{d} \sim R^2_\textrm{c}/D$, and the differences between solute concentrations for different $\Pe_\textrm{s}$ at a given dimensional time $t'$ are apparent. However, these differences almost vanish if the times are normalised by $t'_\textrm{d}$. This is expected for an isotropic diffusive process, thus shown to be the dominant mechanism of mass transfer inside the cell. 

\section{Details of bi-spherical coordinate calculations}\label{bisp_calc}

\begin{figure}[t]
\begin{center}

\includegraphics[width=\linewidth]{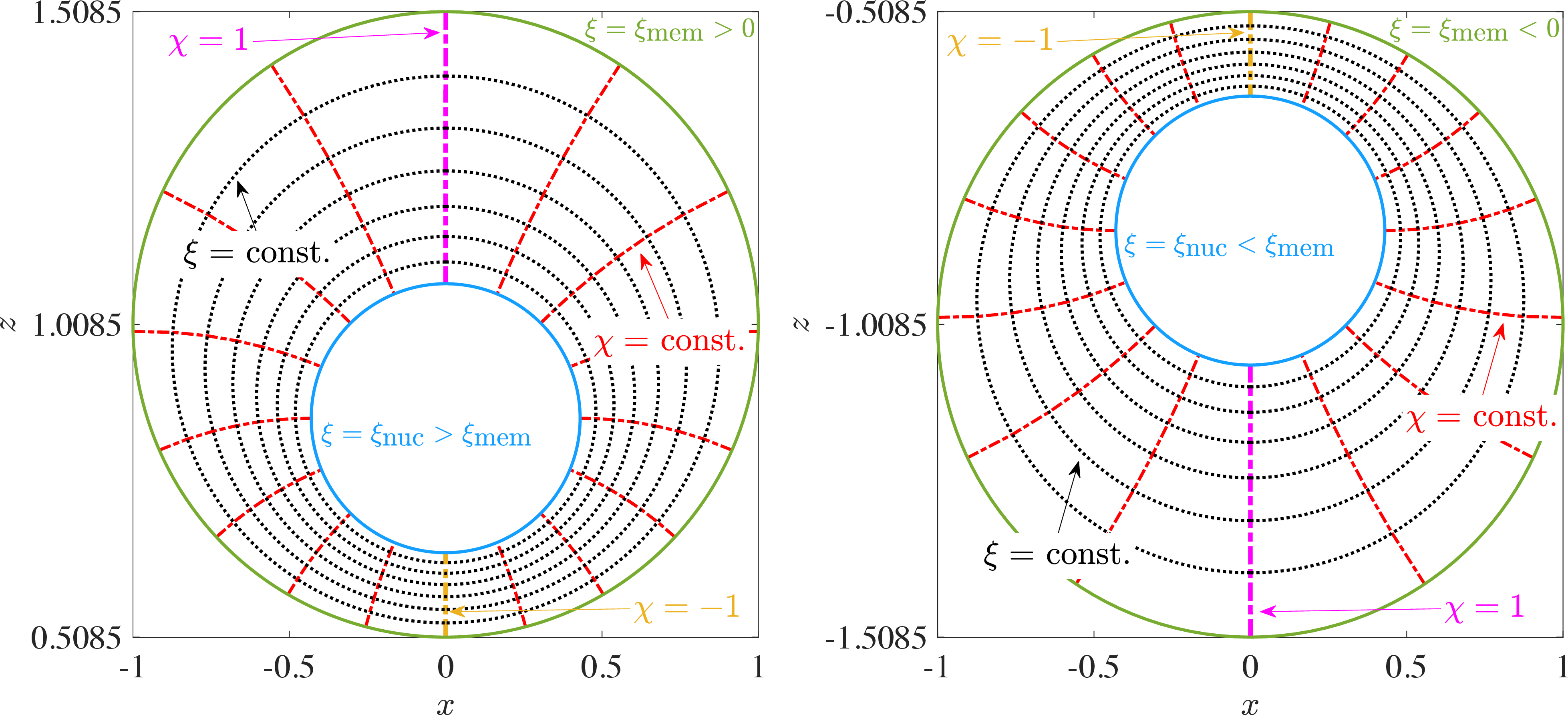}

\caption{Depiction of the bi-spherical coordinate grid for negative (left) and positive (right) vertical eccentricities, with $\left| e_z \right| = 0.3$. The radius ratio between spheres given by $\xi = \xi_\textrm{nuc}$ and $\xi = \xi_\textrm{mem}$ is $\kappa = 0.43$. The velocity components $\left( u_{\xi}, u_{\chi} \right)$ are directed normal to the $\left( \xi, \chi \right)$ iso-surfaces and are positive toward the direction of increasing $\left( \xi, \chi \right)$.}
\label{BiSp_grid_2}
\end{center}
\end{figure}

The geometry of the bi-spherical coordinate system is shown in Fig~\ref{BiSp_grid_2}. The Cartesian coordinates $\left( x, y, z \right)$ are expressed in terms of the bi-spherical coordinates $\left( \xi, \chi, \phi \right)$ as
\begin{align}
    x &= \frac{a \sqrt{1 - \chi^2} }{\cosh \xi - \chi}\cos \phi, \\ \nonumber
    y &= \frac{a \sqrt{1 - \chi^2} }{\cosh \xi - \chi}\sin \phi, \\ \nonumber
    z &= \frac{a \sinh \xi}{\cosh \xi - \chi},
\end{align}
where $a = \left| \sinh \left( \xi_\textrm{mem} \right) \right|$. The bi-spherical coordinates enable us to represent the nuclear and cell membranes as surfaces with a constant value of the $\xi$ coordinate. These values at the nucleus $\left( \xi_\textrm{nuc} \right)$ and the membrane surface $\left( \xi_\textrm{mem} \right)$ are functions of the eccentricity $e_z$ and the radius ratio $\kappa$,
\begin{align}\label{xi_vals_e_kappa}
    \xi_\textrm{mem} &= -\frac{e_z}{\left| e_z \right|}\cosh ^{-1} \left( \frac{1 - \kappa + e_z^2}{2 \left| e_z \right|} \right), \\ \nonumber
    \xi_\textrm{nuc} &= -\frac{e_z}{\left| e_z \right|}\cosh ^{-1} \left( \frac{1 - \kappa - e_z^2}{2 \kappa \left| e_z \right|} \right).
\end{align}
Note that $\left| \xi_\textrm{nuc} \right|$ is always greater than $\left| \xi_\textrm{mem} \right|$, and that both $\xi_\textrm{nuc}$ and $\xi_\textrm{mem}$ are positive for $e_z<0$ and negative for $e_z>0$. With this  geometrical setup, we may now describe the solution of eqns.~\eqref{Stokes_natural_conv_ND} and \eqref{lap_T_ND} in bi-spherical coordinates.

\subsection{Temperature}

The general solution to Laplace's equation for the normalised temperature field, $\nabla^2 \Theta = 0$, in bi-spherical coordinates is given by
\begin{equation}\label{genSoln_Lap_BiSph}
    \Theta(\xi, \chi) = \left( \cosh \xi - \chi \right)^{1/2} \sum_{n=0}^{\infty} { \Theta_n(\xi) L_n(\chi)},
\end{equation}
where $L_n(\chi)$ are the Legendre polynomials, which are solutions of the ordinary differential equation
\begin{equation}\label{Leg_eqn}
    \frac{d}{d \chi} \left\{ \left( 1 - \chi^2 \right) \frac{d}{d \chi} L_n\left( \chi \right) \right\} = -n\left( n + 1 \right)L_n\left( \chi \right),
\end{equation}
the functions $\Theta_n$, called temperature modes, are given by
\begin{equation}\label{theta_n}
    \Theta_n(\xi) = b_n \sinh \left\{ (n+1/2) \left| \xi \right| \right\} + d_n \cosh \left\{ (n+1/2)\xi \right\}.
\end{equation}
In practice, one truncates the summation in eqn.~\eqref{genSoln_Lap_BiSph} to a finite number of terms, beyond which additional terms cause negligible change in the distribution $\Theta \left( \xi, \chi \right)$ (i.e.~the solution has converged). In the present work, we truncate the summation once the maximum relative error caused by adding an extra term falls below 0.1\%. The constants $\left\{ b_n, d_n \right\}$ in eqn.~\eqref{theta_n} are found by applying the boundary conditions $\Theta(\xi_\textrm{nuc},\chi) = 1$ and $\Theta(\xi_\textrm{mem},\chi) = 0$, which yields
\begin{align}\label{an_bn}
    b_n &= \frac{ \sqrt{2} e^{-(n+1/2)\left| \xi_\textrm{nuc} \right|} \cosh \left\{ (n+1/2)\xi_\textrm{mem} \right\} }{ \sinh \left\{ (n+1/2)\left( \left| \xi_\textrm{nuc} \right| - \left| \xi_\textrm{mem} \right| \right) \right\} }, \\ \nonumber
    d_n &= -\frac{ \sqrt{2} e^{-(n+1/2) \left| \xi_\textrm{nuc} \right|} \sinh \left\{ (n+1/2)\left| \xi_\textrm{mem} \right| \right\} }{ \sinh \left\{ (n+1/2) \left( \left| \xi_\textrm{nuc} \right| - \left| \xi_\textrm{mem} \right| \right) \right\} }.
\end{align}
The temperature problem is therefore completely solved and it provides the driving force for the  fluid flow via the Stokes eqns.~\eqref{Stokes_natural_conv_ND}, which we solve in the next section.

\subsection{Flow field}

The flow field is axisymmetric and is expressed in terms of the Stokes flow streamfunction, $\Psi^\textrm{b} \left( \xi, \chi \right)$, as
\begin{align}\label{vel_BiSp}
    \ub &= u_{\xi} \mathbf{i}_\xi + u_{\chi} \mathbf{i}_\chi, \\ \nonumber
    &= \frac{\left( \cosh \xi - \chi \right)^2}{a^2} \left( \frac{\partial \Psi^\textrm{b}}{\partial \chi} \mathbf{i}_{\xi} - \frac{1}{\sqrt{1-\chi^2}} \frac{\partial \Psi^\textrm{b}}{\partial \xi} \mathbf{i}_{\chi} \right),
\end{align}
where the super-script `b' denotes the coordinates (bi-spherical) in which the streamfunction is defined. The velocity components $\left( u_\xi, u_\chi \right)$ are directed normal to the $\left( \xi, \chi \right)$ iso-surfaces (see Fig~\ref{BiSp_grid_2}) and are positive in the direction of increasing $\left( \xi, \chi \right)$. The unit vectors $\left( \mathbf{i}_\xi, \mathbf{i}_\chi \right)$ are expressed in terms of the Cartesian coordinate unit vectors $\left( \mathbf{i}_x, \mathbf{i}_y, \mathbf{i}_z \right)$ as
\begin{align}\label{unit_vec_conv_BiSp}
    \mathbf{i}_\xi &= \frac{1 - \chi \cosh \xi}{\cosh \xi - \chi} \mathbf{i}_z - \frac{\sqrt{1 - \chi^2} \sinh \xi}{\cosh \xi - \chi} \left( \mathbf{i}_x \cos \phi + \mathbf{i}_y \sin \phi \right), \\ \nonumber
    \mathbf{i}_\chi &= \frac{\sqrt{1 - \chi^2} \sinh \xi}{\cosh \xi - \chi} \mathbf{i}_z + \frac{1 - \chi \cosh \xi}{\cosh \xi - \chi} \left( \mathbf{i}_x \cos \phi + \mathbf{i}_y \sin \phi \right).
\end{align}
The equation governing the streamfunction can be derived from the Stokes eqns.~\eqref{Stokes_natural_conv_ND} as~\cite{HappelBrenner2012, Leal2007}
\begin{equation}\label{streamFunc_eqn_bisp}
    -\frac{\cosh \xi - \chi}{a \sqrt{1 - \chi^2}} E^2 \left[ E^2 \left( \Psi^\textrm{b} \right) \right] \left( -\mathbf{i}_x \sin \phi + \mathbf{i}_y \cos \phi \right) = \nabla \times \left( \Theta \mathbf{i}_z \right),
\end{equation}
where $E^2$ is the differential operator
\begin{equation}\label{E2_bisp}
    E^2 \left( \Psi^\textrm{b} \right) \equiv \frac{\cosh \xi - \chi}{a^2} \left[ \frac{\partial}{\partial \xi} \left( \left( \cosh \xi - \chi \right) \frac{\partial \Psi^\textrm{b}}{\partial \xi} \right) + (1-\chi^2)\frac{\partial}{\partial \chi} \left( \left( \cosh \xi - \chi \right) \frac{\partial \Psi^\textrm{b}}{\partial \chi} \right) \right].
\end{equation}
The general solution to eqn.~\eqref{streamFunc_eqn_bisp} requires us to write $\Psi^\textrm{b}$ as
\begin{equation}\label{Psi_b_general}
    \Psi^\textrm{b}(\xi, \chi) = \left( \cosh \xi - \chi \right)^{-3/2} \sum_{n=0}^{\infty} { (1-\chi^2) \frac{dL_n}{d \chi} U_n(\xi) },    
\end{equation}
where $U_n(\xi)$ are unknown functions (velocity modes), that will depend linearly on the temperature modes $\Theta_n$ (eqn.~\eqref{theta_n}). If the summations in eqns.~\eqref{genSoln_Lap_BiSph} and \eqref{Psi_b_general} are truncated after $N$ terms, then we need to solve for $N$ functions $U_n(\xi)$, $1 \le n \le N$. Substituting eqns.~\eqref{Psi_b_general} and \eqref{genSoln_Lap_BiSph} into eqn.~\eqref{streamFunc_eqn_bisp} yields
\begin{align}\label{gde_Psi_pre_proj_BS}
    \sum_{n=0}^{\infty} { -\Gamma^3 \frac{2(n+1)}{a^4} \frac{L_{n+1}(\chi) - \chi L_n(\chi)}{1-\chi^2} \mathcal{E}^\textrm{b}_n(\xi) } &= \sum_{p=0}^{\infty} {\left( 1 - \chi \cosh \xi  \right) \frac{dL_p}{d \chi} \Theta_p(\xi) } \\ \nonumber &- \sum_{p=0}^{\infty} {\frac{L_p(\chi)}{2} \left( \Theta_p(\xi) \cosh \xi + 2 \frac{d \Theta_p}{d \xi} \sinh \xi \right) },
\end{align}
where $\Gamma \equiv \left( \cosh \xi - \chi \right)$, and $\mathcal{E}^\textrm{b}_n(\xi)$ is the  function
\begin{equation}\label{Un_poly}
    \mathcal{E}^\textrm{b}_n(\xi) = \frac{1}{2} \frac{d^4 U_n}{d \xi^4} - \left( n^2 + n + \frac{5}{4} \right)\frac{d^2 U_n}{d \xi^2} + \left( \frac{n^4}{2} + n^3 - \frac{n^2}{4} - \frac{3n}{4} + \frac{9}{32} \right) U_n(\xi).
\end{equation}
We next multiply eqn.~\eqref{gde_Psi_pre_proj_BS} by $(1-\chi^2)({dL_i}/{d\chi})$ and integrate over the limits $\chi=-1$ to $\chi=1$. When this projection is carried out for $1 \le i \le N$, it yields a system of $N$ coupled, linear ordinary differential equations for the velocity functions $\left\{ U_1(\xi), U_2(\xi),\;..., U_N(\xi) \right\}$. These functions are expressible in terms of the  {known} modal distribution of the temperature $\Theta_p(\xi)$. The projection onto $(1-\chi^2)dL_i/d\chi$ of the right-hand-side of eqn.~\eqref{gde_Psi_pre_proj_BS} can be simplified using the properties of Legendre polynomials as
\begin{equation}\label{RHS_BiSp_projected_3}
    \frac{2i(i+1)}{2i+1} \left[ \Theta_i(\xi) - \frac{\cosh \xi}{2}\left\{ \Theta_{i+1}(\xi) + \Theta_{i-1}(\xi)\right\} - \sinh \xi \left\{ \frac{d \Theta_{i - 1}/d \xi}{2i-1} - \frac{d \Theta_{i + 1}/d \xi}{2i+3} \right\}\right].
\end{equation}

The projection onto $(1-\chi^2)dL_i/d\chi$ of the left-hand-side of eqn.~\eqref{gde_Psi_pre_proj_BS} yields
\begin{equation}\label{LHS_BiSp_projected}
    -\frac{2}{a^4} \sum_{n=0}^{\infty} {\int_{-1}^{1} { \left( \cosh \xi - \chi \right)^3 (n+1) \left\{ L_{n+1}(\chi) - \chi L_{n}(\chi) \right\} \frac{dL_i}{d \chi} \mathcal{E}^\textrm{b}_n(\xi) } \,d\chi},
\end{equation}
which can be written in short-hand as
\begin{equation}\label{LHS_BiSp_projected_short}
    -\frac{2}{a^4} \sum_{n=0}^{\infty} { \mathcal{I}_{in}(\xi) \mathcal{E}^\textrm{b}_n(\xi) },
\end{equation}
where
\begin{align}\label{I_in_xi}
    \mathcal{I}_{in}(\xi) &= \int_{-1}^{1} { \left( \cosh \xi - \chi \right)^3 (n+1) \left\{ L_{n+1}(\chi) - \chi L_{n}(\chi) \right\} \frac{dL_i}{d \chi} \,d\chi}\\ \nonumber
    &= \int_{-1}^{1} { i(n+1) \left( \cosh \xi - \chi \right)^3 \left\{ L_{n+1}(\chi) - \chi L_{n}(\chi) \right\} \frac{ \chi L_{i}(\chi) - L_{i-1}(\chi) }{\chi^2 - 1} \,d\chi}.
\end{align}
The second line of eqn.~\eqref{I_in_xi} is obtained by making use of the following recurrence relation for Legendre polynomials,
\begin{equation*}
    \frac{1 - \chi^2}{i}\frac{dL_i}{d \chi} = L_{i-1}(\chi) - \chi L_{i}(\chi).
\end{equation*}
In left-hand-side of eqn.~\eqref{I_in_xi}, the index `$i$' in $\mathcal{I}_{in}$ denotes the term onto which the projection is made: $(1-\chi^2)dL_i/d\chi$, and the index `$n$' denotes the contribution to the projection from the $n^\textrm{th}$ term in the expansion on the left-hand-side of eqn.~\eqref{gde_Psi_pre_proj_BS}. 

The next step is to write out~\eqref{LHS_BiSp_projected_short} for a discrete number of points, say $M$, along $\xi$. Thus, each mode $U_n(\xi)$ is evaluated on $M$ points: $\left\{ U_n(\xi_1), U_n(\xi_2),\;...,U_n(\xi_M) \right\}$, where $\xi_1=\xi_\textrm{nuc}$ and $\xi_M=\xi_\textrm{mem}$. The discrete version of eqn.~\eqref{LHS_BiSp_projected_short} at $\xi=\xi_l\;\left( 1 \le l \le M \right)$ reads
\begin{equation}\label{LHS_BiSp_projected_short_2}
    -\frac{2}{a^4} \sum_{n=0}^{\infty} { \mathscr{I}_{lin} \mathcal{E}^\textrm{b}_n(\xi_l) },
\end{equation}
where $\mathscr{I}_{lin}$ is a three-dimensional matrix resulting from {the evaluation of $\mathcal{I}_{in}(\xi)$ at $\xi = \xi_l$}, i.e.~$\mathscr{I}_{lin} = \mathcal{I}_{in}(\xi=\xi_l)$. If $N$ is the number of modes (or the upper limit of the summation) at which the temperature (eqn.~\eqref{genSoln_Lap_BiSph}) and the streamfunction (eqn.~\eqref{Psi_b_general}) expansions are truncated, then the size of $\mathscr{I}_{lin}$ is $M \times N \times N$. At this level of modal resolution/truncation, eqn.~\eqref{gde_Psi_pre_proj_BS} written for $\xi = \xi_l$ and projected onto $\left( 1 - \chi^2 \right) \frac{dL_i}{d \chi}$ is thus given by
\begin{align}\label{eqn_BiSp_projected_short}
    -\frac{2}{a^4} \sum_{n=0}^{N} { \mathscr{I}_{lin} \mathcal{E}^\textrm{b}_n(\xi_l) } &= \frac{2i(i+1)}{2i+1} \left[ \Theta_i(\xi_l) - \frac{\cosh \xi_l }{2}\left\{ \Theta_{i+1}(\xi_l) + \Theta_{i-1}(\xi_l)\right\} \right] \\ \nonumber
    & - \frac{2i(i+1)}{2i+1} \left[ \sinh \xi_l \left. \left( \frac{d \Theta_{i-1}/ d \xi}{2i-1} - \frac{d \Theta_{i+1}/ d \xi}{2i+3} \right) \right|_{\xi=\xi_l} \right].
\end{align}
While the right-hand-side of eqn.~\eqref{eqn_BiSp_projected_short} is explicitly known (see eqn.~\eqref{theta_n}), the derivatives of the functions $U_n(\xi)$ (inherent in definition of $\mathcal{E}^\textrm{b}_n(\xi)$; see eqn.~\eqref{Un_poly}) in the left-hand-side need to be discretised. In the present work, we use second-order accurate finite differences for this discretisation. This leads to a total of $N \times M$ unknowns, $U_n(\xi_l)$, where $1 \le n \le N$ and $1 \le l \le M$.

\begin{figure}[t]
\begin{center}

\includegraphics[width=13cm]{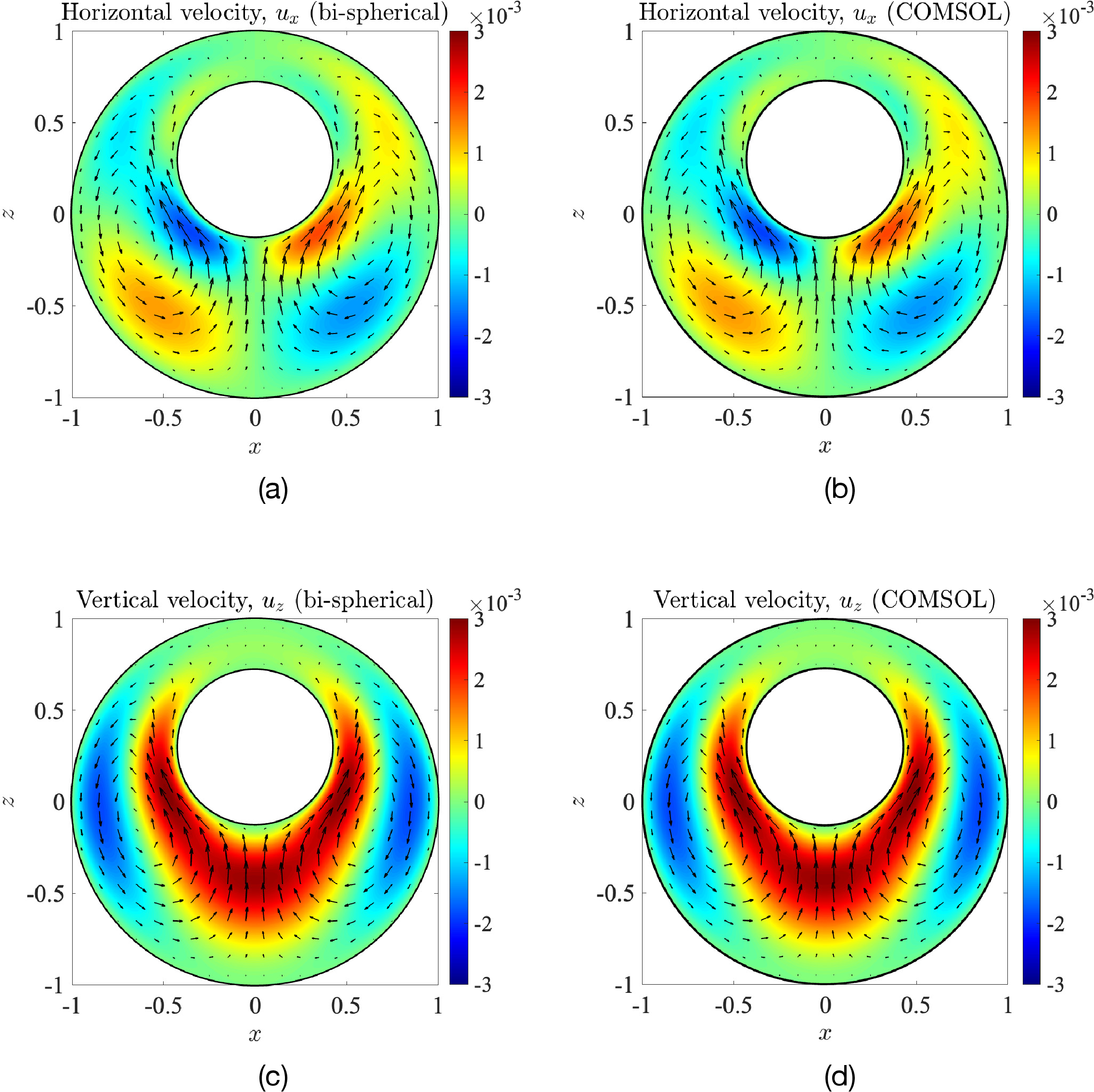}

\caption{Comparison of the  dimensionless horizontal 
components of the flow field at the $y=0$ mid-plane
(panels (a) and (b)) and vertical components (panels (c) and (d)), calculated using the bi-spherical coordinates methodology (left column) and via COMSOL simulations (right column). The geometrical parameters are given by $\kappa = 0.43, e_x = 0, e_z = 0.3$.}

\label{BiSp_vs_COMSOL}
\end{center}
\end{figure}

\subsection{Boundary conditions for the flow}

Eqn.~\eqref{eqn_BiSp_projected_short} is a fourth-order, linear, coupled ordinary differential equation for the functions $U_n(\xi)$. It is supplemented by four boundary conditions involving velocity components $\left( u_{\xi}, u_{\chi} \right)$ at the inner and outer sphere surfaces (i.e.~at the nucleus and the cell membrane),
\begin{align}\label{BCs_BiSp}
    u_\xi \left( \xi = \xi_\textrm{nuc} \right) & = u_\xi \left( \xi = \xi_\textrm{mem} \right) = 0, \\ \nonumber
    u_\chi \left( \xi = \xi_\textrm{nuc} \right) & = u_\chi \left( \xi = \xi_\textrm{mem} \right) = 0.
\end{align}
Using eqn.~\eqref{vel_BiSp}, one can show that these boundary conditions yield the following equations for $U_n(\xi)$ and $dU_n/d \xi$,
\begin{align}\label{BCs_BiSp_2}
    U_n(\xi_\textrm{nuc}) &= U_n(\xi_\textrm{mem}) = 0, \\ \nonumber
    \left. \frac{dU_n}{d \xi} \right|_{\xi = \xi_\textrm{nuc}} &= \left. \frac{dU_n}{d \xi} \right|_{\xi = \xi_\textrm{mem}} = 0.
\end{align}
Eqns.~\eqref{eqn_BiSp_projected_short} and \eqref{BCs_BiSp_2} provide the $N \times M$ linear equations required to obtain the unknown functions $U_n(\xi_l)$. Inverting this system of equations solves the hydrodynamic problem for the flow field $\textbf{u}(\xi,\chi)$ (obtained via eqns.~\eqref{vel_BiSp} and \eqref{Psi_b_general}). A transformation from $\textbf{u}(\xi,\chi)$ to $\textbf{u}(x,y,z)$ (using eqn.~\eqref{unit_vec_conv_BiSp}) allows us to plot and compare the velocities obtained using the above methodology with the finite-element COMSOL simulations. This comparison is shown in Fig~\ref{BiSp_vs_COMSOL} where we obtain essentially an identical match between the two solutions. Note that in bi-spherical coordinates, we have solved the problem in a grid where the $z$-coordinate of the cell centre is not given by $z=0$ (see Fig~\ref{BiSp_grid_2}), but while plotting our results to compare with those from COMSOL simulations, we have shifted the domain along the $z$-axis such that the cell centre always lies at $\left( x=0, z=0 \right)$.

\section{Details of analytical solution in the concentric case}\label{sph_calc}

\begin{figure}[t]
\begin{center}
\includegraphics[width=6cm]{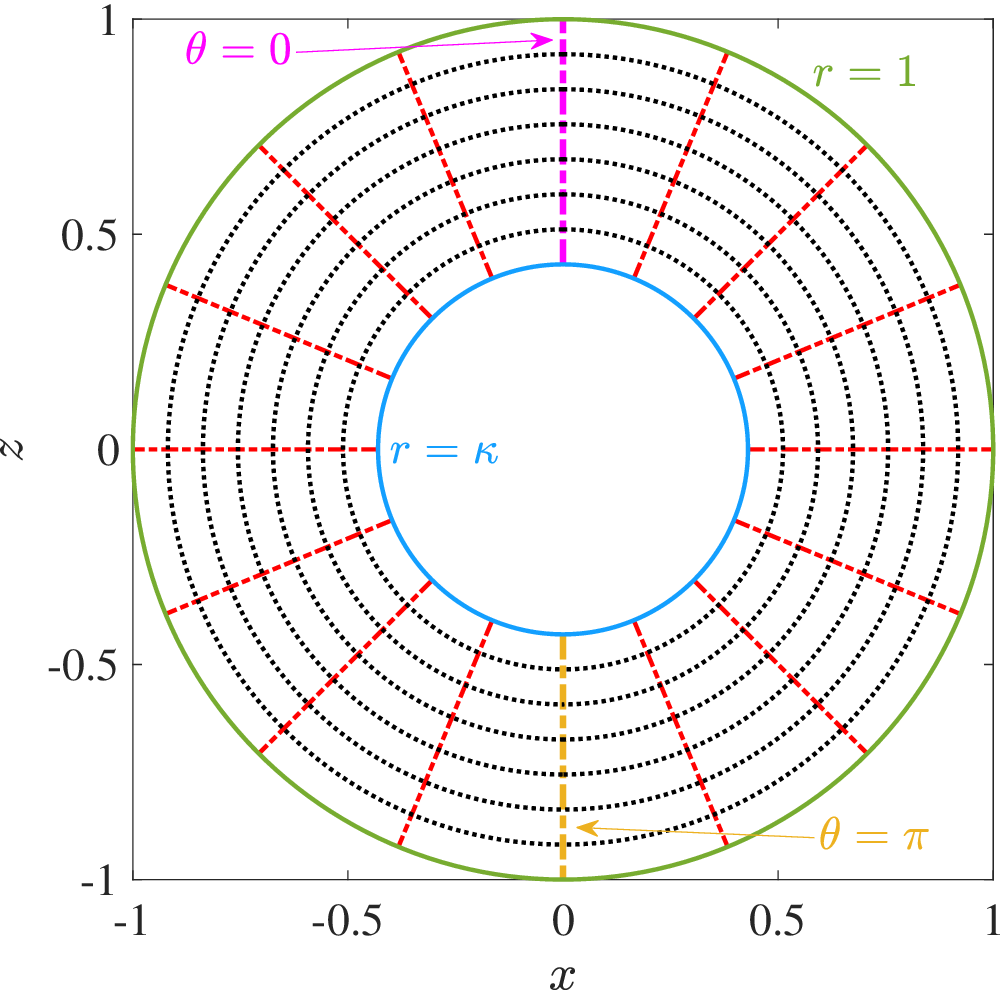}

\caption{The axisymmetric spherical coordinates, with $\kappa \le r \le 1$ and $0 \le \theta \le \pi$. The velocity components $\left( u_r, u_{\theta} \right)$ are directed normal to the $\left( r, \theta \right)$ iso-surfaces and are positive toward the direction of increasing $\left( r, \theta \right)$.}

\label{Sp_coords}
\end{center}
\end{figure}

In the concentric case, the nucleus and cell have the same centre. This geometry is much simpler to handle than the axisymmetric (but eccentric) arrangement of the previous section, which required bi-spherical coordinates (Appendix~\ref{bisp_calc}). As a result, one can solve the problem using spherical polar coordinates and obtain fully analytical solutions for the temperature and flow fields.  {While this problem has been solved in Ref.~\cite{Mack1968} in the context of natural convection between concentric spheres at low thermal P\'eclet numbers, we repeat here the solution in our notation, for the sake of completeness. We obtain the same solution as the leading-order solution in Ref.~\cite{Mack1968}, once the notations are made consistent.} The problem geometry is shown in Fig~\ref{Sp_coords}. The Cartesian coordinates $\left( x, y, z \right)$ are expressed in terms of the spherical polar coordinates $\left( r, \theta, \phi \right)$ as
\begin{align}
    x &= r \sin \theta \cos \phi, \\ \nonumber
    y &= r \sin \theta \sin \phi, \\ \nonumber
    z &= r \cos \theta.
\end{align}

\subsection{Temperature}

The thermal problem is driven solely by diffusion because the thermal diffusivity $\alpha$ is large enough to give vanishing thermal P\'eclet number, $\Pe_\textrm{t} = u_\textrm{ref} R_\textrm{c}/ \alpha \ll 1$ (see also Fig~\ref{vel_contours_AD_vs_oD}). Since the boundary conditions (see eqns.~\eqref{T_BCs_ND}) are independent of the polar angle $\theta$, the normalised temperature field, $\Theta$, must also be independent of $\theta$, i.e.~it is spherically symmetric. The normalised temperature $\Theta$ must then satisfy Laplace's equation, given by
\begin{equation}\label{theta_diff}
    \frac{1}{r^2}\frac{d}{dr}\left( r^2 \frac{d \Theta}{dr}\right) = 0,
\end{equation}
along with the boundary conditions of prescribed temperature on the nucleus and membrane,
\begin{align}\label{T_BCs_ND_2}
    \Theta \left( r=\kappa \right) &= 1, \\ \nonumber
    \Theta \left( r=1 \right) &= 0.
\end{align}
The solution to eqns.~\eqref{theta_diff} and \eqref{T_BCs_ND_2} is classically given by 
\begin{equation}\label{theta_r}
    \Theta(r) = \frac{\kappa}{1 - \kappa}\left( \frac{1}{r} - 1 \right).
\end{equation}

\subsection{Flow field}
Here also, the flow can be represented in terms of the streamfunction for Stokes flow, $\Psi^\textrm{s} \left( r, \theta \right)$,
\begin{align}\label{ur_uth}
    \ub &= u_r \mathbf{i}_r + u_{\theta} \mathbf{i}_{\theta}, \\ \nonumber
        &= \frac{1}{r \sin \theta} \left( -\frac{1}{r}\frac{\partial \Psi^\textrm{s}}{\partial \theta} \mathbf{i}_r + \frac{\partial \Psi^\textrm{s}}{\partial r} \mathbf{i}_{\theta} \right),
\end{align}
where the super-script `s' now denotes that the streamfunction has been defined in terms of spherical coordinates. The spherical basis vectors $\left( \mathbf{i}_r, \mathbf{i}_{\theta} \right)$ are expressed in terms of the Cartesian coordinate unit vectors $\left( \mathbf{i}_x, \mathbf{i}_y, \mathbf{i}_z \right)$ as
\begin{align}\label{unit_vec_conv_Sp}
    \mathbf{i}_r &= \sin \theta \cos \phi \mathbf{i}_x + \sin \theta \sin \phi \mathbf{i}_y + \cos \theta \mathbf{i}_z, \\ \nonumber
    \mathbf{i}_{\theta} &= \cos \theta \cos \phi \mathbf{i}_x + \cos \theta \sin \phi \mathbf{i}_y - \sin \theta \mathbf{i}_z.
\end{align}
In spherical coordinates, the streamfunction satisfies
\begin{equation}\label{streamFunc_eqn_sp}
    -\frac{1}{r \sin \theta}E^2 \left[ E^2 \left( \Psi^\textrm{s} \right) \right] = \frac{\sin \theta}{r^2}\frac{\kappa}{1-\kappa};
\end{equation}
the differential operator $E^2$ is given by
\begin{equation}\label{E2_sp}
    E^2 \left( \Psi^\textrm{s} \right) = \frac{\partial^2 \Psi^\textrm{s}}{\partial r^2} + \frac{1-\mu^2}{r^2} \frac{\partial^2 \Psi^\textrm{s}}{\partial \mu^2},
\end{equation}
where we write $\mu \equiv \cos \theta$. The general solution for $\Psi^\textrm{s}$ is
\begin{equation}\label{Psi_general_sph}
    \Psi^\textrm{s}(r, \mu) = \sum_{n=0}^{\infty} { (1-\mu^2) \frac{d L_n}{d \mu} f_n(r) },
\end{equation}
where the $L_n(\mu)$ are the Legendre polynomials as defined in eqn.~\eqref{Leg_eqn}. The structure of eqn.~\eqref{streamFunc_eqn_sp} allows us to pose an Ansatz for $\Psi^\textrm{s}(r, \mu)$: the angular dependence in that equation can only be balanced if we restrict the expansion \eqref{Psi_general_sph} to just the $n=1$ term, i.e.
\begin{equation}
    \Psi^\textrm{s}(r, \theta) = f_1(r) \sin^2 \theta.
\end{equation}
 Substitution into eqn.~\eqref{streamFunc_eqn_sp} then gives the following governing equation for $f_1(r)$,
\begin{equation}\label{gde_f1r}
    \frac{d^4 f_1}{d r^4} + 2 \left\{ -\frac{2}{r^2}\frac{d^2 f_1}{d r^2} + \frac{4}{r^3} \frac{d f_1}{d r} - \frac{4}{r^4} f_1(r) \right\} = -\frac{1}{r} \frac{\kappa}{1-\kappa}.
\end{equation}

\subsection{Boundary conditions for the flow}

The fourth-order ordinary differential equation \eqref{gde_f1r} is supplemented by the boundary conditions at the nucleus $\left( r = \kappa \right)$ and the cell membrane $\left( r = 1 \right)$. Noting that they have been modelled as rigid spheres, the normal and tangential velocities at these surfaces must vanish,
\begin{align}\label{BCs_Sp_1}
    u_r \left( r = \kappa \right) &= u_r \left( r = 1 \right) = 0, \\ \nonumber
    u_{\theta} \left( r = \kappa \right) &= u_{\theta} \left( r = 1 \right) = 0,
\end{align}
which convert to the following conditions on the function $f_1(r)$,
\begin{align}\label{BCs_Sp_2}
    f_1(r = \kappa) &= f_1(r = 1) = 0, \\ \nonumber
    \left. \frac{df_1}{dr} \right|_{r=\kappa} &= \left. \frac{df_1}{dr} \right|_{r=1} = 0.
\end{align}
The full solution to eqns.~\eqref{gde_f1r} and \eqref{BCs_Sp_2} is provided in eqns.~\eqref{f1r} and \eqref{c_i_f1r} in the main text. Additionally, the vertical and horizontal velocities plotted in Fig~\ref{Sph_vs_COMSOL} are obtained by transforming the spherical coordinate velocity representation $\ub \left( r, \theta \right)$ to a Cartesian representation $\ub \left( x, y, z \right)$ as
\begin{align}\label{uz_ux}
u_z\left( r, \theta \right) & = -\left( \frac{2 \cos^2 \theta}{r^2}f_1\left(r\right) + \frac{\sin^2 \theta}{r} \frac{df_1}{dr} \right), \\ \nonumber
u_x\left( r, \theta, \phi \right) & = \left( -\frac{1}{r^2}f_1\left(r\right) + \frac{1}{2r}\frac{df_1}{dr} \right) \sin \left( 2 \theta \right) \cos \phi.
\end{align}

\nolinenumbers

%%% contents of .bbl file: START

%%% contents of .bbl file: STOP

\end{document}